\documentclass[showpacs,showkeys,twocolumn,nofootinbib]{revtex4}

\usepackage{amsmath}
\usepackage{amssymb}
\usepackage{slashed}
\usepackage[dvips]{graphicx}
\usepackage{color}

\usepackage{verbatim} 
\usepackage[sans]{dsfont}

\renewcommand{\d}{\ensuremath{\mathrm{d}}}
\newcommand{\hc}{\ensuremath{\mathrm{h.c.}}}
\newcommand{\C}{\ensuremath{c}}

\newcommand{\SU}[1]{\ensuremath{\mathrm{SU}(#1)}}

\newcommand{\U}[1]{\ensuremath{\mathrm{U}(#1)}}

\renewcommand{\d}{\ensuremath{\mathrm{d}}}

\begin{document}
\def\im{\mathrm{i}}
\def \openone {\leavevmode\hbox{\small1\kern-3.0pt\normalsize1}}
\def\beginm{\left(\begin{array}}
\def\endm{\end{array}\right)}

\title{Top-quark and neutrino composite Higgs bosons}

\author{Adam Smetana}
\email{Adam.Smetana@utef.cvut.cz}
\affiliation{Institute of Experimental and Applied Physics, Czech Technical University in Prague, Horsk\'{a} 3a/22, 128 00 Prague 2, Czech Republic}

\begin{abstract}
In the context of top-quark condensation models, the top-quark alone is too light to saturate the correct value of the electroweak scale by its condensate. Within the seesaw scenario the neutrinos can have their Dirac masses large enough so that their condensates can provide significant contribution to the value of the electroweak scale. We address the question of a phenomenological feasibility of the top-quark and neutrino condensation conspiracy against the electroweak symmetry. Mandatory is to reproduce the masses of electroweak gauge bosons, the top-quark mass and the recently observed scalar mass of $125\,\mathrm{GeV}$ and to satisfy the upper limits on absolute value of active neutrino masses. To accomplish that we design a reasonably simplified effective model with two composite Higgs doublets. Additionally, we work with a general number $N$ of right-handed neutrino flavor triplets participating on the seesaw mechanism. There are no experimental constraints limiting this number. The upper limit is set by the model itself. Provided that the condensation scale is of order $10^{17-18}\,\mathrm{GeV}$ and the number of right-handed neutrinos is ${\cal O}(100-1000)$, the model predicts masses of additional Higgs bosons below $250\,\mathrm{GeV}$ and a suppression of the top-quark Yukawa coupling to the $125\,\mathrm{GeV}$ particle at the $\sim60\,\%$ level of the Standard model value.
\end{abstract}



\maketitle

\section{Introduction}
\label{intro}

The Standard model reproduces the measured data surprisingly well. There is only little room for beyond-Standard model physics. The recently measured value for the Higgs boson mass $M_h\doteq125\,\mathrm{GeV}$ \cite{125Higgs:2012gk,125Higgs:2012gu} however leads to the running quartic coupling constant $\lambda$ vanishing at the scale below the Planck scale \cite{Degrassi:2012ry,Bednyakov:2013eba,Chetyrkin:2013wya}. 
There is a conceivable interpretation: The Higgs boson reveals its compositeness at that scale. It corresponds to the historical experience that all of the observed scalar particles ultimately turned out to be fermion bound states.

The mechanism of the electroweak symmetry breaking manifests itself by the fact that both the top-quark mass $m_t$ and the Higgs boson mass $M_h$ are of the same order as the electroweak scale $v$ which is the scale for mass of weak bosons, e.g., $M_W=g_2 v/2$. A simple contemplation leads to the suspicion that both the longitudinal components of the weak vector bosons and the Higgs boson are in fact bound states of top-quark. This is the idea of top-quark condensation models introduced in \cite{Hosek:1985jr,Miransky:1988xi,Bardeen:1989ds}. The key idea is that to break the electroweak symmetry it is enough to generate dynamically masses of known fermions, in particular of the top-quark. In this sense the top-quark mass generation is primary. The electroweak symmetry breaking comes out automatically. The top-quark mass then determines the magnitude of the electroweak scale and the Higgs boson mass.

Although the actual calculation gives a correct order of magnitude of masses, there are two essential failures of the simplest model when comparing it with experiment. First, the top-quark is observed to be too light to saturate the electroweak scale $v$. Keeping the top-quark condensation scale below the Planck scale, the top-quark condensation alone can provide only at most 68\,\% of the $W$ and $Z$ boson masses. Second, the Higgs boson is predicted to be too heavy, in all available calculations $M_h>m_t$. This prediction was ruled out already before the actual measurement of $125\,\mathrm{GeV}$ particle at the LHC \cite{125Higgs:2012gk,125Higgs:2012gu}.

If we want to maintain the attractive idea that the top-quark condensation is the source of the electroweak symmetry breaking, we need to improve the simplest model.

First, we need to suppose yet another source potent to saturate the value of the electroweak scale. The remaining observed quarks and charged leptons bring no improvement in this respect as they are way too light and contribute by truly negligible amount to the electroweak scale. At this point one could be easily seduced to invoke some new yet unobserved fermions with high enough mass, like, for instance, fourth-generation fermions or techniquarks. This however may not be necessary. Among the known fermions, there is an additional source of the electroweak symmetry breaking naturally present in the form of the neutrino Dirac mass $m_D$ provided that the seesaw mechanism is at work. If the neutrino Dirac mass is of the order of the electroweak scale, $m_D\sim v$, then the neutrino condensate is strong enough to complement the electroweak scale \cite{Martin:1991xw,Cvetic:1992ps,Antusch:2002xh}.

Second, once we identify two main fermion sources of the electroweak symmetry breaking we resort to the reasonably simplified description using correspondingly two composite Higgs doublets. The presence of more than single Higgs doublet is instrumental in achieving lighter neutral scalar as the candidate for the $125\,\mathrm{GeV}$ particle due to the appropriate mixing among components of the doublets. 

The idea of top-quark and neutrino condensation was addressed already in the past. First, Martin \cite{Martin:1991xw} investigated a model in which the idea was implemented in the simplest possible way. He invoked a factorization assumption on four-fermion interactions which resulted in the low-energy description with only single Higgs doublet. He reached the correct value of the top-quark mass, but from now-days perspective, the model suffers from exhibiting too heavy Higgs boson particle, in the same way as the original top-quark-alone condensation models \cite{Miransky:1988xi,Bardeen:1989ds}. Ten years later, the issue was addressed again by Antusch et al. confirming its usefulness for obtaining the correct value of the top-quark mass and suggesting that the two-Higgs-doublet low-energy description is worthwhile to study in more detail. Another ten years later we are addressing the idea once again and confront it with the new experimental evidence of the $125\,\mathrm{GeV}$ boson excitation.

We believe that analyzing the top-quark and neutrino condensation is relevant for the class of models where all types of known fermions condense \cite{Hosek:1982cz,Kimura:1984zv,Nagoshi:1990wk,Cvetic:1992xn,Gribov:1994jy,Bashford:2003yg,Brauner:2004kg,Benes:2008ir,Wetterich:2006ii,Schwindt:2008gj,Hosek:2009ys,Hosek:NagoyaProceeding,Benes:2011gi}. The models share the underlying principle of primary role of known fermion masses in the electroweak symmetry breaking. Those models which provide the seesaw mechanism for neutrino masses, like for instance \cite{Cvetic:1992ps,Smetana:2011tj}, have a chance to escape from being ruled out by the measurement of the top-quark mass, just because of the presence of the second potentially big Dirac mass $m_D$. Currently, however, they have to be confronted with the measurement of $125\,\mathrm{GeV}$ particle. We do not refer to any of these underlying models in our analysis, we rather emulate their dynamics by particular four-fermion interaction. Effectively, instead of using a multitude of composite Higgs doublets, one for each Dirac mass, we work with only two Higgs doublets, one for each significant source of the electroweak symmetry breaking.

The number of right-handed neutrino types participating on the seesaw mechanism is not constrained by any upper limit \cite{Ellis:2007wz,Heeck:2012fw}. As claimed there, higher number of the order ${\cal O}(100)$ is even well motivated within some string constructions. Large number of right-handed neutrinos ${\cal O}(100)$ has also an improving effect on the standard thermal leptogenesis \cite{Eisele:2007ws}. Being of order ${\cal O}(10-100)$ it can serve as the reason for large lepton mixing angles \cite{Feldstein:2011ck}. Therefore we study the dependence of our results on the number of right-handed neutrinos.

The precise form of the neutrino mass spectrum is not known. For our analysis of the electroweak symmetry breaking it does not play an essential role. Therefore we simulate the unknown mass spectrum of neutrinos by the most simple choice for the neutrino mass matrix, which is characterized by a common Dirac mass $m_D$, by a common right-handed Majorana mass $M_R$ and by the number $N$ of right-handed neutrino flavor triplets. By this simplification we can control the order of magnitude of active neutrino masses but do not reproduce the neutrino physics exactly. That would otherwise require to specify the underlying dynamics in detail.

In this work we explore the possibility to saturate the electroweak symmetry breaking by combined sources from both top-quark and neutrino Dirac condensates.
We concentrate on the analysis of the low-energy mass spectrum, mainly to the possibility to achieve one of the composite scalar as light as $125\,\mathrm{GeV}$ while reproducing correct values for the masses of the electroweak gauge bosons and the top-quark and satisfying the upper limits on absolute value of active neutrino masses $m_\nu<0.2\,\mathrm{eV}$. We study the coupling properties of the lighter Higgs scalar to the top-quark and gauge bosons. Next we calculate the mass spectrum of additional Higgs bosons. We study the sensitivity of the results on the number $N$ of right-handed neutrino flavor triplets. 

In the Sect.~\ref{SecII} we introduce the lagrangian for our analysis and identify relevant symmetries. In the Sect.~\ref{SecIII} we formulate the effective description using two Higgs doublets and write the low-energy lagrangian. In the Sect.~\ref{SecIV} we study the mass spectrum of the model as a function of parameters of the low-energy lagrangian. In the Sect.~\ref{SecV} we formulate the renormalization group equations which govern the evolution of the low-energy parameters from the condensation scale down to the electroweak scale. In the Sect.~\ref{SecVI} we present results. In the Sect.~\ref{SecVII} we discuss the results and confront them with experimental data. Finally, in the Sect.~\ref{SecVII} we conclude.

\section{Underlying four-fermion interaction}\label{SecII}

\subsection{Underlying lagrangian}
\label{sec:II.1}

For the purpose of our analysis we let only the top-quark and the neutrinos to condense and to contribute to the electroweak scale. Therefore we define our simplified model by the four-fermion interaction
\begin{equation}\label{4f}
{\cal L}_{4f} = - G_t(\bar t_Rq_L)(\bar q_Lt_R) - G_\nu(\sum_s\bar \nu_{Rs}\ell_L)(\sum_{s'}\bar \ell_L\nu_{Rs'}) \,
\end{equation}
which is designed to provide us just by the condensation of the desired form discussed in the introduction. It reflects the main assumption that only the top-quark and neutrinos play the appreciable role in what we address in our work - the electroweak symmetry breaking.

Only the third generation of quarks participates in the interaction. The fields $t_R$ and $q_L$ are color triplets. On the other hand, because we suppose that all neutrino Dirac masses are of the order of electroweak scale then within the simplified model we are letting all three generations of leptons to participate in the interaction. Therefore the fields $\nu_R$ and $\ell_L$ are flavor triplets. The left-handed fields are weak isospin doublets. All three types of indices are suppressed. The explicitly written index $s=1,\dots,N$ labels $N$ right-handed neutrino flavor triplets. By this simplified dynamics we are going to generate masses of only top-quark and neutrinos.

If the underlying dynamics is such that the four-fermion interactions follow from an exchange of neutral and colorless gauge bosons, then there are only these two effective terms relevant for the top-quark and neutrino condensation. No mixing terms like $\propto(\bar t_Rq_L)(\bar \ell_L\nu_{R})$ appear. There could appear also various four-fermion interactions of other leptons and quarks, but we neglect them here as they play rather negligible role in the electroweak symmetry breaking.

For the sake of simplicity we use here a factorization assumption on the neutrino coupling constants, so that there is single neutrino coupling constant $G_\nu$ and single neutrino-Higgs boson doublet, in the same spirit as in \cite{Martin:1991xw}. After the condensation, this provides a degenerate Dirac mass for neutrinos.

We introduce the right-handed neutrino Majorana mass term within their kinetic lagrangian
\begin{equation}\label{L_M_R}
{\cal L}_{\nu_R} = \bar{\nu}_{Rs}\im\slashed{\partial}\nu_{Rs}-\frac{1}{2}M_R\overline{\nu_{Rs}^\C}\nu_{Rs}+\hc \,.
\end{equation}
We take the Majorana mass degenerate and diagonal for the same sake of simplicity.
We dare to accept these simplifications as our aim is not to reproduce the neutrino phenomenology exactly.

For our purpose we assume the lagrangian
\begin{eqnarray}
{\cal L} & = & {\cal L}_{\mathrm{usual}}+{\cal L}_{\mathrm{model}} \,, \\
{\cal L}_{\mathrm{model}} & = & {\cal L}_{4f}+{\cal L}_{\nu_R} \,,
\end{eqnarray}
where ${\cal L}_{\mathrm{usual}}$ contains kinetic terms of all known fermions, their Standard model gauge interactions and pure gauge boson terms.

\subsection{Symmetries}
\label{sec:II.2}

The lagrangian ${\cal L}_{\mathrm{model}}$ has well separated quark and lepton sectors. On the classical level, it is invariant under the global symmetry\footnote{The rest of standard fermions and their corresponding symmetries are of course present in the model in order to provide proper  anomaly cancelation, but we hold them back here as they do not participate in the symmetry breaking in our simplified analysis. Due to the factorization assumption the three generations of leptons exhibit a single common symmetry group.}
\begin{eqnarray}
G_{\mathrm{model}} &=& \big[\SU{2}\times\U{1}^2\big]_q\times\big[\SU{2}\times\U{1}\big]_\ell \,. \label{doubled_symm}
\end{eqnarray}
One subgroup of $G_{\mathrm{model}}$ is the electroweak $\SU{2}_L\times\U{1}_Y$ gauge symmetry group. The electroweak interactions explicitly break the symmetry $G_{\mathrm{model}}$, so the symmetry of the full lagrangian ${\cal L}$ is
\begin{equation}\label{EW_breaking}
G=\SU{2}_L\times\U{1}_Y\times\U{1}_{B}\times\U{1}_{X} \,,
\end{equation}
among which $\SU{2}_L\times\U{1}_Y$ are the weak isospin and weak hypercharge gauge symmetries, $\U{1}_B$ is the baryon number and $\U{1}_X$ is the axial symmetry
\begin{equation}\label{U1X}
\U{1}_X:\ \ X(q_L,t_R,\ell_L,\nu_R)=(-1,0,1,0) \,.
\end{equation}
We are making this choice of $X$ charges in order to have $X(\bar t_Rq_L)=-1$ and $X(\bar \nu_R\ell_L)=+1$. It is this symmetry which prevents the top-quark and neutrino sectors from mixing.

On the quantum level, the group $\U{1}_{X}$ has the axial anomaly. The axial anomaly is provided by the electroweak and QCD dynamics. Additionally, it can be provided by some new not specified dynamics underlying the four-fermion interaction \eqref{4f} like, e.g., the gauge flavor dynamics \cite{Benes:2011gi}. In the following, we will simply parameterize the strength of the anomaly by the scale $\mu_{t\nu}$.

The dynamically generated Dirac masses for top-quark and neutrinos break spontaneously the $G_{\mathrm{model}}$ symmetry down to
\begin{equation}\label{Gmodel_H}
G_{\mathrm{model}} \stackrel{m_t,m_\nu}{\longrightarrow} \big[\U{1}^2\big]_q\times\big[\U{1}\big]_\ell \,.
\end{equation}
It would give rise to 6 massless Nambu--Goldstone bosons.

There are however effects of both the electroweak dynamics and of the axial anomaly which eliminate the 6 massless states completely.
The electroweak interactions change the spontaneous symmetry breaking pattern to
\begin{equation}\label{G_H}
G \stackrel{m_t,m_\nu}{\longrightarrow} \U{1}_\mathrm{em}\times\U{1}_{B} \,.
\end{equation}
Therefore three of the Nambu--Goldstone states are eaten by the electroweak gauge bosons. The other two form a single charged pseudo-Nambu--Goldstone particles whose mass results from the explicit breaking by the electroweak dynamics \eqref{EW_breaking} and it is therefore proportional to the electroweak gauge coupling constants. The remaining single state stays massless if we neglect the effect of the $\U{1}_{X}$ axial anomaly, otherwise it is the pseudo-Nambu--Goldstone boson with the mass proportional to the scale $\mu_{t\nu}$.

\section{Two Higgs doublet description}\label{SecIII}

Effectively, the top-quark and neutrino condensation can be described by the condensation of two composite Higgs doublets
\begin{eqnarray}
H_t & \sim & (\bar t_Rq_L)\,, \\
H_\nu & \sim & (\sum_s\bar \nu_{Rs}\ell_L) \,.
\end{eqnarray}
Using them we can rewrite the four-fermion interaction \eqref{4f} via the Hubbard--Stratonovich transformation \cite{Stratonovich:1957aa,Hubbard:1959aa} as
\begin{eqnarray}\label{boson_lagr}
{\cal L}_{4f} & = & - y_{t0}(\bar q_Lt_R)H_t - y_{\nu0}(\sum_s\bar \ell_L\nu_{Rs})H_\nu+\hc \nonumber\\
& & +\mu_{t0}^{2}H_{t}^\dag H_{t}+\mu_{\nu0}^{2}H_{\nu}^\dag H_{\nu} \,,
\end{eqnarray}
which is completely equivalent to \eqref{4f} as far as $H_t$ and $H_\nu$ are non-propagating auxiliary fields.

Below the condensation scale $\Lambda$, we substitute the original lagrangian for the effective lagrangian formed by operators allowed by the symmetries and generated by radiative corrections. Among the operators, there are kinetic terms for the composite Higgs doublets $H_{t}$ and $H_{\nu}$. If we take only the renormalizable operators we are effectively obtaining two-Higgs-doublet model.
\begin{eqnarray}\label{eff_lagrangian}
{\cal L}_\mathrm{eff} & = & |D H_{t}|^2 + |D H_{\nu}|^2 - {\cal V}(H_t,H_\nu) \nonumber\\
& & - y_t(\bar q_Lt_R)H_t - y_\nu(\sum_s\bar \ell_L\nu_{Rs})H_\nu + \hc \,.
\end{eqnarray}
The potential for the two Higgs doublets invariant with respect to the $G$ symmetry is
\begin{eqnarray}
{\cal V} & = & {\cal V}_{0}+{\cal V}_{\mathrm{EW}}+{\cal V}_{\mathrm{soft}} \\
{\cal V}_{0} & = & -\mu_{t}^2H_{t}^\dag H_{t}-\mu_{\nu}^2H_{\nu}^\dag H_{\nu} \nonumber \\
& & +\tfrac{1}{2}\lambda_t(H_{t}^\dag H_{t})^2+\tfrac{1}{2}\lambda_\nu(H_{\nu}^\dag H_{\nu})^2 \nonumber \\
{\cal V}_{\mathrm{EW}} & = & \lambda_{t\nu}(H_{t}^\dag H_{t})(H_{\nu}^\dag H_{\nu}) +\lambda_{t\nu}'(H_{t}^\dag H_{\nu})(H_{\nu}^\dag H_{t}) \nonumber \,.
\end{eqnarray}
We sort the terms in the potential for Higgs bosons according to their primary origin. Those terms denoted by ${\cal V}_{0}$ are generated due to the four-fermion interaction irrespectively of the presence of the electroweak dynamics which provides only corrections to their magnitude. The terms denoted by ${\cal V}_{\mathrm{EW}}$ on the other hand make a bridge between the top-quark and neutrino sectors generated only because of the presence of the electroweak dynamics. They vanish in the limit of vanishing electroweak coupling constants.

In order to take into account the axial anomaly of $\U{1}_X$ we introduce additional term which mixes the two Higgs doublets and breaks explicitly the $\U{1}_X$ symmetry
\begin{eqnarray}
{\cal V}_{\mathrm{soft}} & = & -\mu_{t\nu}^2H_{t}^\dag H_{\nu}+\hc \,.
\end{eqnarray}
This term can not be generated at any loop order neither by the four-fermion interaction nor by the electroweak dynamics. In this work we use $\mu_{t\nu}$ as a free parameter.

Apart from $\mu_{t\nu}$ all the parameters of the lagrangian ${\cal L}_\mathrm{eff}$, i.e., $y$'s, $\mu$'s, and $\lambda$'s, run with the renormalization scale $\mu$ according to the renormalization group equations towards the condensation scale $\mu=\Lambda$. At the condensation scale they are linked to the values of the underlying lagrangian ${\cal L}$ via the field renormalization factors. Because of its non-perturbative origin, the mixing parameter $\mu_{t\nu}$ is not the subject of the renormalization group equations and as such it acts as a free parameter in our analysis.

The quartic stability of the potential is given by the conditions
\begin{eqnarray}
&&\lambda_t,\ \lambda_\nu>0\,, \\
&&\sqrt{\lambda_t\lambda_\nu}>-\lambda_{t\nu}-\lambda_{t\nu}' \,.
\end{eqnarray}
The parameter setting in the range
\begin{equation}
\lambda_{t\nu}'<0\ \ \mathrm{and}\ \ \mu_{t\nu}^2>0
\end{equation}
leads to the minimum of the potential which conserves electric charge. This is completely analogous to the work \cite{Luty:1990bg} in the context of top-quark and bottom-quark two-Higgs-doublet model.

\section{Masses}\label{SecIV}

\subsection{Electroweak scale and fermion masses}
\label{sec:IV.1}

The top-quark and the neutrino contributions to the electroweak symmetry breaking are given by the condensates $v_t\propto\langle\bar tt\rangle$, and $v_\nu\propto\langle\bar \nu\nu\rangle$.

From the lagrangians \eqref{L_M_R} and \eqref{eff_lagrangian} the neutrino mass matrix $\mathbf{M}_\nu$ turns out to be
\begin{equation}\label{nu_mass_matrix}
\mathbf{M}_\nu = \beginm{cccc} 0 & \mathbf{m}_D & \cdots & \mathbf{m}_D \\ \mathbf{m}_D & \mathbf{m}_R & 0 &  0 \\
\vdots & 0 & \ddots & 0 \\  \mathbf{m}_D & 0 & 0 & \mathbf{m}_R  \endm \,,
\end{equation}
where $\mathbf{m}_D\equiv \tfrac{1}{\sqrt{2}}y_\nu v_\nu\openone$ and $\mathbf{m}_R\equiv M_R\openone$, where $\openone$ is the $3\times3$ unit matrix in the flavor space.
Each left-handed neutrino therefore mixes with $N$ right-handed neutrinos, but do not mix among each other.

We assume that both $v_t$ and $v_\nu$ together saturate the electroweak scale $v$,
\begin{equation}\label{ew_scale}
v^2=v_{t}^2+v_{\nu}^2 \,.
\end{equation}
This is the main idea of this work.

Next we define $\beta$-angle by
\begin{equation}\label{beta}
\tan\beta\equiv\frac{v_{t}}{v_{\nu}} \,.
\end{equation}
We choose the convention that $\beta\in\langle0,\tfrac{\pi}{2}\rangle$.

The mass of the top-quark is given by equation
\begin{equation}\label{top_mass}
m_{t}=y_{t}(\mu=m_t)v_{t}/\sqrt{2} \,
\end{equation}
and the light neutrino mass is given as a smaller eigenvalue of the neutrino mass matrix \eqref{nu_mass_matrix} by the seesaw formula
\begin{equation}\label{nu_mass}
m_{\nu}=\frac{N y_{\nu}^2(\mu=m_\nu)v_{\nu}^2}{2M_R} \,,
\end{equation}
where $y_{t,\nu}(\mu)$ are running Yukawa coupling constants.

Within the two-composite-Higgs-doublet model the values of Yukawa couplings around the electroweak scale are calculable using their renormalization group evolution down from the condensation scale $\Lambda$. Because $\Lambda$ is very large the couplings are only weakly sensitive to their initial values $y_{t,\nu}(\Lambda)$ as they have enough time to approach an infrared fixed point.

The equation \eqref{top_mass} for the top-quark mass, $m_t\doteq172\,\mathrm{GeV}$, then fixes $v_t$. From the relation \eqref{ew_scale} we determine $v_\nu$ -- a portion of the electroweak scale left for neutrinos. Finally from \eqref{nu_mass} we determine the right-handed neutrino Majorana mass $M_R$ based on the assumption that $m_\nu\lesssim0.2\,\mathrm{eV}$. Having $|\mathbf{m}_D|\sim v$ implies roughly $M_R\gtrsim10^{14}\,\mathrm{GeV}$. Here the construction closes by an important restriction on the right-handed neutrino Majorana mass and the condensation scale
\begin{equation}\label{non_decoupling_cond}
\Lambda_\mathrm{Planck}>\Lambda>M_R\,.
\end{equation}
If the Majorana mass $M_R$ were too big, then the correspondingly heavy right-handed neutrinos would decouple from the dynamics \cite{Martin:1991xw} before they would manage to condense with the left-handed neutrinos. Finally, we find the top-quark and neutrino condensation meaningful only if the condensation scale $\Lambda$ is below the Planck scale.

\subsection{Higgs boson masses}
\label{sec:IV.2}

After the electroweak symmetry is broken by the condensate $v^2=v_{t}^2+v_{\nu}^2$, we get Higgs boson mass eigenstates from ${\cal L}_\mathrm{eff}$ in the unitary gauge in terms of two neutral scalars $\phi^{0}_t$ and $\phi^{0}_\nu$, one neutral pseudo-scalar $A$ and one charged scalar $H^\pm$,
\begin{subequations}\label{Htnu_redef}
\begin{eqnarray}
H_t & = & \beginm{c}  \tfrac{1}{\sqrt{2}}(v_t+\phi^{0}_t+\im\cos\beta A) \\ -\im\cos\beta H^-  \endm \,, \\
H_\nu & = & \beginm{c} \tfrac{1}{\sqrt{2}}(v_\nu+\phi^{0}_\nu-\im\sin\beta A) \\ \im\sin\beta H^- \endm \,.
\end{eqnarray}
\end{subequations}
The quadratic lagrangian contains a mixing of the neutral scalars $\phi^{0}_t$ and $\phi^{0}_\nu$. Its diagonalization results in the mass eigenstates $h$ and $H$ characterized by the mixing angle $\alpha$ according to
\begin{eqnarray}
H & = & \sqrt{2}\big[\phi^{0}_{t}\sin\alpha+\phi^{0}_{\nu}\cos\alpha\big] \,,\\
h & = & \sqrt{2}\big[\phi^{0}_{t}\cos\alpha-\phi^{0}_{\nu}\sin\alpha\big] \,,\\
\tan2\alpha & = & \frac{(\lambda_{t\nu}+\lambda_{t\nu}')v_t v_\nu - \mu_{t\nu}^2}{-\tfrac{1}{2}(v_{t}^2\lambda_t-v_{\nu}^2\lambda_\nu)-\mu_{t\nu}^2\cot2\beta} \,.
\end{eqnarray}
We choose the convention that $\alpha\in\langle-\tfrac{\pi}{2},0)$. In this case the lighter Higgs scalar is always $h$.
The masses for the neutral scalars $H$ and $h$, for the neutral pseudo-scalar $A$ and for the charged scalar $H^\pm$ are given by
\begin{subequations}\label{Higgs_masses}
\begin{eqnarray}
M_{H,h}^2 & = & \frac{1}{2}f_{\pm}(t=\ln M_{H,h}) \,, \\
M_{A}^2 & = & \frac{2\mu_{t\nu}^2}{\sin2\beta} \,, \\
M_{H^\pm}^2 & = & \frac{2\mu_{t\nu}^2}{\sin2\beta}-\frac{1}{2}\lambda_{t\nu}'(t=\ln M_{H^\pm})v^2
\end{eqnarray}
\end{subequations}
where
\begin{eqnarray}
f_{\pm}(t) & = &  v_{t}^2\lambda_t(t)+v_{\nu}^2\lambda_\nu(t)+\frac{2\mu_{t\nu}^2}{\sin2\beta}\pm\sqrt{A(t)} \,, \nonumber\\
A(t)       & = &  \big(v_{t}^2\lambda_t(t)-v_{\nu}^2\lambda_\nu(t)\big)^2+4v_{t}^2v_{\nu}^2\big(\lambda_{t\nu}(t)+\lambda_{t\nu}'(t)\big)^2 \nonumber\\
               & & +\frac{4\mu_{t\nu}^4}{\sin^22\beta}-2\mu_{t\nu}^2\big[v_{t}^2\lambda_t(t)\tan\beta+v_{\nu}^2\lambda_\nu(t)\cot\beta \nonumber\\
             &   & -v_tv_\nu\big(\lambda_t(t)+\lambda_\nu(t)+4\lambda_{t\nu}(t)+4\lambda_{t\nu}'(t)\big)\big] \,.\nonumber
\end{eqnarray}
We have introduced the logarithmic renormalization scale $t\equiv\ln\mu$.

\begin{figure}[t]
\resizebox{0.5\textwidth}{!}{
\includegraphics{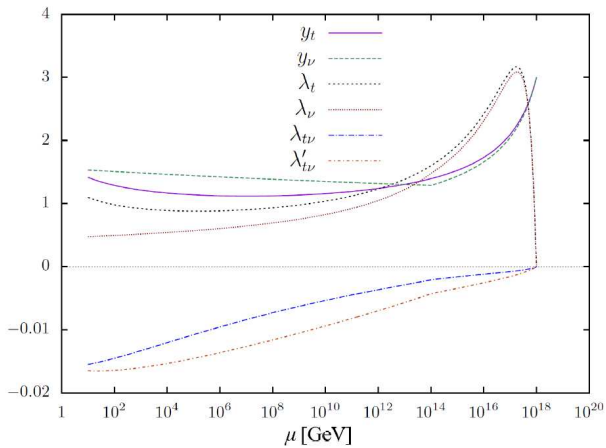}
}
\caption{ \small The renormalization group evolution of $y_t$, $y_\nu$, $\lambda_t$, $\lambda_\nu$, $\lambda_{t\nu}$ and $\lambda_{t\nu}'$ for the parameter setting \eqref{param_set}. The role of $M_R=10^{14}\,\mathrm{GeV}$ is visible as the threshold in the evolution. }
\label{plot_RGE}
\end{figure}

In the following we will show that our model leads to
\begin{eqnarray}\label{cond_1}
\lambda_{t\nu},\ \lambda_{t\nu}'<0
\end{eqnarray}
and
\begin{equation}\label{cond_2}
\lambda_t,\ \lambda_\nu\gg|\lambda_{t\nu}|,\ |\lambda_{t\nu}'| \,.
\end{equation}
For illustration, see Fig.~\ref{plot_RGE}.
In order to conserve electric charge we will investigate only the values
\begin{equation}
\mu_{t\nu}^2\geq0\,.
\end{equation}
Let us investigate two important limits:

1) First, by setting $\lambda_{t\nu},\lambda_{t\nu}'\rightarrow0$, we switch the electroweak interactions off, and by setting $\mu_{t\nu}^2=0$, we preserve the $\U{1}_X$ symmetry. The top-quark and neutrino condensation causes the symmetry breaking \eqref{G_H}. In this limit the spectrum of bosons changes to
\begin{eqnarray}
& M_{A}^2=M_{H^\pm}^2 = 0\,, & \\
& M_{H}^2=\mathrm{Max}\big\{v_{t}^2\lambda_t,v_{\nu}^2\lambda_\nu\big\}\,,& \\
& M_{h}^2=\mathrm{Min}\big\{v_{t}^2\lambda_t,v_{\nu}^2\lambda_\nu\big\} & \label{MH2_mu_0} \,,
\end{eqnarray}
where we identify three uneaten Nambu--Goldstone bosons. The two Higgs scalars do not mix as $\alpha=-\tfrac{\pi}{2}$. The identity of the lighter Higgs scalar is given by
\begin{eqnarray}
\mathrm{for}\ \ \ v_{t}^2\lambda_t>v_{\nu}^2\lambda_\nu & : & \ \ \ h=\sqrt{2}\phi^{0}_{\nu}\,, \\
\mathrm{for}\ \ \ v_{t}^2\lambda_t<v_{\nu}^2\lambda_\nu & : & \ \ \ h=\sqrt{2}\phi^{0}_{t}\,.
\end{eqnarray}

2) Second, we set $\lambda_{t\nu},\lambda_{t\nu}'\rightarrow0$ again, but we let $\mu_{t\nu}\gg v$.
In this limit, the spectrum of bosons changes to
\begin{eqnarray}
& M_{H}^2 = M_{A}^2=M_{H^\pm}^2 = \frac{2\mu_{t\nu}^2}{\sin2\beta} \,, & \\
& M_{h}^2 = \tfrac{1}{8}v^2\big[4\lambda_t(\sin^2\beta+\sin^4\beta)+4\lambda_\nu(\cos^2\beta+\cos^4\beta) & \nonumber\\
&  
-(\lambda_t+\lambda_\nu)\sin^22\beta\big]\,.  & \label{MH2_mu_inf}
\end{eqnarray}
In this limit four degrees of freedom $H$, $A$ and $H^\pm$ get degenerate masses proportional to $\mu_{t\nu}$ and decouple from the low-energy physics. One degree of freedom $h$ stays light. It is the mixture of top-quark and neutrino neutral composite scalars which is characterized by $\tan2\alpha=\tan2\beta$, hence the mixing angle $\alpha=\beta-\tfrac{\pi}{2}$.

To obtain values of masses we need to evolve the running parameters from their initial values at the condensation scale down to the electroweak scale. This we will do in the next section.

\subsection{Interactions of mass eigenstates}
\label{sec:IV.3}

We list here several interactions of the lighter Higgs scalar $h$ and of the charged Higgs boson $H^\pm$, which are particulary important for today phenomenological analysis:
\begin{subequations}
\begin{eqnarray}
{\cal L}_{\mathrm{Yukawa}}^{h} & = & -\frac{m_t}{v}C_th\bar tt \,, \label{Higgs_Yukawa}\\
{\cal L}_{\mathrm{gauge}}^{h} & = & -C_Vh\left(\frac{2M_{W}^2}{v}W^+W^- + \frac{M_{Z}^2}{v}Z^2\right) \,, \label{Higgs_gauge}\\
{\cal L}_{\mathrm{H^\pm}}^{h} & = & -C_{H^\pm}vhH^+H^-\,, \label{charged_Higgs_Higgs}
\end{eqnarray}
\end{subequations}
where the scaling coupling factors are
\begin{subequations}\label{coupling_parameters}
\begin{eqnarray}
C_t & = & \frac{\cos\alpha}{\sin\beta} \,, \\
C_V & = & \sin(\beta-\alpha) \,, \\
C_{H^\pm} & = & \sin\beta\cos\alpha\big(\sin^2\beta\lambda_{t\nu}+\cos^2\beta(\lambda_t-\lambda_{t\nu}')  \big) \\
            & & -\cos\beta\sin\alpha\big(\cos^2\beta\lambda_{t\nu}+\sin^2\beta(\lambda_\nu-\lambda_{t\nu}')\big) \nonumber\,.
\end{eqnarray}
\end{subequations}
They measure the departure from the Standard model, which is characterized by $C_t=1$, $C_V=1$ and $C_{H^\pm}=0$.

\section{Renormalization group equations}\label{SecV}

The parameters of the low-energy lagrangian ${\cal L}_\mathrm{eff}$ run with the renormalization scale $\mu$ according to the equations of renormalization group. The exhausting analysis of renormalization group equations for two-Higgs-doublet models is to be found in \cite{Hill:1985tg}. The compositeness of the Higgs doublets is expressed by the fact that the lagrangian \eqref{eff_lagrangian} is equivalent to the Higgs-less lagrangian \eqref{4f} or \eqref{boson_lagr} at the condensation scale $\Lambda$. From sewing the two lagrangian together at $\Lambda$, a set of boundary conditions for $\mu\rightarrow\Lambda$ follows
\begin{eqnarray}
& y_t \rightarrow \infty\,,\ \ y_\nu \rightarrow \infty\,, \ \ y_t/y_\nu \rightarrow y_{t0}/y_{\nu0} , & \nonumber\\
& \lambda_t/y_{t}^4 \rightarrow 0\,,\ \ \lambda_\nu/y_{\nu}^4 \rightarrow 0\,, & \nonumber\\
& \lambda_{t\nu}/y_{t}^2y_{\nu}^2 \rightarrow 0\,,\ \ \lambda_{t\nu}'/y_{t}^2y_{\nu}^2 \rightarrow 0\,, & \nonumber\\
& \mu_{t}^2/y_{t}^2 \rightarrow \mu_{t0}^{2}/y_{t0}^{2}\,,\ \ \mu_{\nu}^2/y_{\nu}^2 \rightarrow \mu_{\nu0}^{2}/y_{\nu0}^{2}\,, &
\end{eqnarray}
In practice, for actual numerical calculation, we will use the boundary conditions
\begin{eqnarray}\label{boundary_conditions}
y_t(\ln\Lambda)=Y_t \,, \ \ y_\nu(\ln\Lambda)=Y_\nu \,, \nonumber\\
\lambda_t(\ln\Lambda)=0 \,, \ \
\lambda_\nu(\ln\Lambda)=0 \,, \nonumber\\
\lambda_{t\nu}(\ln\Lambda)=0 \,, \ \
\lambda_{t\nu}'(\ln\Lambda)=0 \,,
\end{eqnarray}
where $Y_t$ and $Y_\nu$ are finite numbers on which the low-energy result depends only very weakly.

Further, we will restrict our analysis only to one loop order.

The presence of the second Higgs doublet affects the $t$ evolution of the gauge coupling constants governed by the one-loop renormalization group equations and the boundary conditions given by the experimental values at $\mu=M_Z$:
\begin{subequations}\label{g123}
\begin{eqnarray}
16\pi^2\frac{\d}{\d t}g_1 = \hphantom{-}7g_{1}^3 \,,&\ \ &g_{1}^2(\ln M_Z)\doteq0.127 \,, \\
16\pi^2\frac{\d}{\d t}g_2 = -3g_{2}^3 \,,&\ \ &g_{2}^2(\ln M_Z)\doteq0.425 \,, \\
16\pi^2\frac{\d}{\d t}g_3 = -7g_{3}^3 \,,&\ \ &g_{3}^2(\ln M_Z)\doteq1.440 \,.
\end{eqnarray}
\end{subequations}
The renormalization group equation for Yukawa coupling constants are
\begin{eqnarray}
16\pi^2\frac{\d}{\d t}y_t   & = & y_t\big[\tfrac{9}{2}y_{t}^2-\tfrac{17}{12}g_{1}^2-\tfrac{9}{4}g_{2}^2-8g_{3}^2\big] \,, \label{Yukawa_RGE} \\
16\pi^2\frac{\d}{\d t}y_\nu & = & y_\nu\big[3(N+\tfrac{1}{2})\theta(t-M_R)y_{\nu}^2-\tfrac{3}{4}g_{1}^2-\tfrac{9}{4}g_{2}^2\big] \,. \nonumber
\end{eqnarray}
The $\theta$-function stands for the threshold below which the heavy right-handed neutrinos decouple from the system. The renormalization group equations for the quartic coupling constants are
\begin{eqnarray}
16\pi^2\frac{\d}{\d t}\lambda_t   & = & 12\lambda_{t}^2+4\lambda_{t\nu}^2+4\lambda_{t\nu}\lambda_{t\nu}'+2{\lambda_{t\nu}'}^2 \nonumber\\
                                  & & +\lambda_{t}\big[12y_{t}^2-3g_{1}^2-9g_{2}^2\big]-12y_{t}^4 \nonumber\\
                                  & & +\tfrac{3}{4}(g_{1}^4-2g_{1}^2g_{2}^2+3g_{2}^4) \,, \nonumber\\
16\pi^2\frac{\d}{\d t}\lambda_\nu & = & 12\lambda_{\nu}^2+4\lambda_{t\nu}^2+4\lambda_{t\nu}\lambda_{t\nu}'+2{\lambda_{t\nu}'}^2 \nonumber\\
                                  & & +\lambda_{\nu}\big[12N\theta(t-M_R)y_{\nu}^2-3g_{1}^2-9g_{2}^2\big] \nonumber\\
                                  & & -12N\theta(t-M_R) y_{\nu}^4+\tfrac{3}{4}(g_{1}^4-2g_{1}^2g_{2}^2+3g_{2}^4) \,, \nonumber\\
16\pi^2\frac{\d}{\d t}\lambda_{t\nu} & = & 2(\lambda_{\nu}+\lambda_{t})(3\lambda_{t\nu}+\lambda_{t\nu}')+4\lambda_{t\nu}^2+2{\lambda_{t\nu}'}^2 \nonumber\\
                                  & &+\lambda_{t\nu}\big[6y_{t}^2+6N\theta(t-M_R)y_{\nu}^2-3g_{1}^2-9g_{2}^2\big] \nonumber\\
                                  & & +\tfrac{3}{4}(g_{1}^4-2g_{1}^2g_{2}^2+3g_{2}^4) \,, \nonumber\\
16\pi^2\frac{\d}{\d t}\lambda_{t\nu}' & = & \lambda_{t\nu}'\big[\lambda_{\nu}+\lambda_{t}+8\lambda_{t\nu}+4{\lambda_{t\nu}'}+6y_{t}^2 \label{quartic_RGE}\\
                                   & & +6N\theta(t-M_R)y_{\nu}^2-3g_{1}^2-9g_{2}^2\big]+3g_{1}^2g_{2}^2 \,. \nonumber
\end{eqnarray}
According to Luty \cite{Luty:1990bg} the one loop renormalization evolution of the dimensionless parameters does not depend on the presence of the mixing parameter $\mu_{t\nu}$.

\begin{figure}[t]
\begin{tabular}{c}
\resizebox{0.5\textwidth}{!}{
\includegraphics{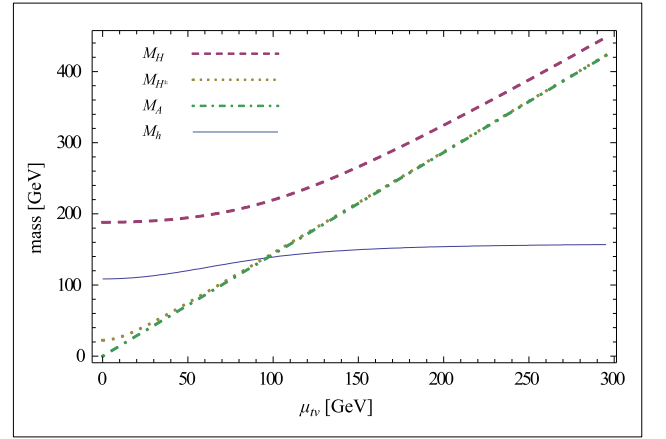}} \\
a) \\
\resizebox{0.5\textwidth}{!}{
\includegraphics{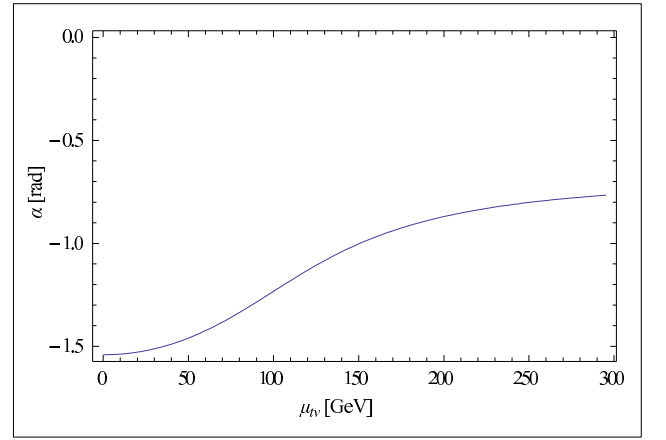}} \\
b)
\end{tabular}
\caption[]{ \small a) Higgs masses $M_{H,h}$, $M_{A}$ and $M_{H^\pm}$ as functions of $\mu_{t\nu}$ for $\Lambda=10^{18}\,\mathrm{GeV}$, $N=1$ and fixed neutrino mass to be $m_\nu=0.2\,\mathrm{eV}$ (i.e. $M_R\sim10^{14}\,\mathrm{GeV}$). b) Mixing of two scalar Higgs bosons $H$ and $h$ as a function of $\mu_{t\nu}$. At the condensation scale, the Yukawa couplings $y_{t,\nu}$ start at the value $Y_{t,\nu}=3$, and the quartic couplings $\lambda$ start at zero value. }
\label{plot_HiggsMasses}
\end{figure}

\section{Results}\label{SecVI}

Before we present our results we briefly describe the strategy of the analysis. Our input parameters are $\Lambda$, $M_R$ and $\mu_{t\nu}$, out of which $M_R$ is fixed by reproducing the neutrino mass $m_\nu=0.2\,\mathrm{eV}$. Strictly speaking, there are additional two parameters $Y_t$ and $Y_\nu$ which however have only mild effect on the result and we take them quite arbitrarily to be
\begin{equation}\label{Yuk_set}
Y_t=Y_\nu=3 \,.
\end{equation}
On top of that we have freedom to choose the number $N$ of right-handed neutrino triplets.

The steps of calculation follow:
1) We solve analytically the equations \eqref{g123} for gauge constants. 2) We numerically evolve the Yukawa and quartic coupling constants according to the equations \eqref{Yukawa_RGE} and \eqref{quartic_RGE} with the boundary conditions \eqref{boundary_conditions} and \eqref{Yuk_set}. 3) From the experimental value of the top-quark mass we determine $v_t$ using the equation \eqref{top_mass}. 4) We calculate $v_\nu$ and $\beta$ from equations \eqref{ew_scale} and \eqref{beta}, respectively. 5) The equation \eqref{nu_mass} gives us the value for the neutrino mass $m_\nu$. Changing the value of $M_R$ we repeat the calculation from the point 2) and iterate the value of the neutrino mass to get $m_\nu=0.2\,\mathrm{eV}$. 6) Using the equations \eqref{Higgs_masses} we calculate the Higss boson mass spectrum.

\subsection{Renormalization group evolution}
\label{sec:VI.1}

The renormalization group evolution of the dimensionless parameters are plotted in Fig.~\ref{plot_RGE} for
\begin{equation}\label{param_set}
N=1,\ \Lambda=10^{18}\,\mathrm{GeV},\ M_R=10^{14}\,\mathrm{GeV}\,.
\end{equation}
As we mentioned before, at one-loop order the result does not depend on the parameter $\mu_{t\nu}$. It gives us a typical result which does not change qualitatively much with changing $\Lambda$ and $M_R$. We can see that at the electroweak scale $\Lambda_\mathrm{EW}$
\begin{eqnarray}
&\lambda_{t\nu},\lambda_{t\nu}'<0 \,,& \\
&100\times\big(|\lambda_{t\nu}|,|\lambda_{t\nu}'|\big)\sim\big(\lambda_{t},\lambda_{\nu}\big) \,, \label{lambda_mag} &
\end{eqnarray}
so the stability conditions \eqref{cond_1} and \eqref{cond_2} are fulfilled. On top of that, taking $\mu_{t\nu}^2>0$ we assure that the condensate will be electrically neutral.

\subsection{Mass spectrum of Higgs bosons}
\label{sec:VI.2}

The typical result \eqref{lambda_mag} allows us to neglect $\lambda_{t\nu}$, $\lambda_{t\nu}'$ in favor of $\lambda_t$, $\lambda_\nu$. That is why the limits analyzed at the end of Sect.~\ref{sec:IV.2} are useful for us. We can roughly estimate the mass of the lighter Higgs scalar $h$ and the mixing angle $\alpha$ to lie in the interval
\begin{equation}
\left.\begin{array}{rcl}M_{h}&\simeq&\langle113,160)\mathrm{GeV} \\
                \alpha&\simeq&\langle-\tfrac{\pi}{2},-0.7)\end{array}\right\}\ \ \mathrm{for}\ \ \mu_{t\nu}=\langle0,\infty)\mathrm{GeV}
\end{equation}
calculated from \eqref{MH2_mu_0} and \eqref{MH2_mu_inf} with input parameters \eqref{param_set} and using estimated values from Fig.~\ref{plot_RGE}, \eqref{ew_scale} and \eqref{top_mass}
\begin{eqnarray}
&v_t\simeq187\,\mathrm{GeV}\,,\ v_\nu\simeq160\,\mathrm{GeV} \,,& \\
&\lambda_t\simeq1.0\,,\ \lambda_\nu\simeq0.5 \,. &
\end{eqnarray}
It represents a promising improvement with respect to the previous results of the single Higgs doublet top-quark condensation models.

Now, let us investigate the solutions of the Higgs boson mass spectrum without approximations. In Fig.~\ref{plot_HiggsMasses}a) we plot the dependence of the Higgs boson masses on the mixing parameter $\mu_{t\nu}$. We use $\Lambda=10^{18}\,\mathrm{GeV}$ while the right-handed neutrino Majorana mass we fix from demanding $m_\nu=0.2\,\mathrm{eV}$. It turns out to be roughly $M_R\sim10^{14}\,\mathrm{GeV}$.

For lower values of $\mu_{t\nu}$ the bosons $A$ and $H^\pm$ are very light. The mass $M_{A}$ even vanishes for vanishing $\mu_{t\nu}$ reflecting the spontaneous breaking of the exact $\U{1}_X$ symmetry \eqref{U1X}. In Fig.~\ref{plot_HiggsMasses}b) where we plot the dependence of the $h$-$H$ mixing angle $\alpha$ on $\mu_{t\nu}$, it can be seen that the lighter scalar $h$ is composed mainly of neutrinos, $\alpha\sim-\tfrac{\pi}{2}$, and therefore its coupling to top-quark is suppressed, see \eqref{Higgs_Yukawa}.

Increasing $\mu_{t\nu}$ translates into the lifting of masses of the Higgs bosons, $H$, $A$ and $H^\pm$. They soon become growing linearly and nearly degenerate. On the other hand, the mass of the lighter scalar $h$ is only mildly sensitive to the increase of $\mu_{t\nu}$ and quite soon saturates just below $160\,\mathrm{GeV}$. On top of that it is acquiring gradually larger admixture from a top-quark composite state, reaching the value over $\alpha\sim-0.8$. $M_{A}$ minimizes the spectrum of $H$, $A$ and $H^\pm$ for all positive values of $\mu_{t\nu}$ in our model.

Setting $\Lambda=10^{18}\,\mathrm{GeV}$, $m_\nu=0.2\,\mathrm{eV}$ and $N=1$ we can reach the lighter Higgs boson mass of the desired value
\begin{equation}
M_{h}=125\,\mathrm{GeV}\ \ \mathrm{for}\ \ \mu_{t\nu}\doteq62\,\mathrm{GeV}\,.
\end{equation}
The value $\mu_{t\nu}\doteq62\,\mathrm{GeV}$ translates into the Higgs boson mass spectrum
\begin{equation}
M_{H}\doteq198\,\mathrm{GeV}\,,\ \ M_{A}\doteq88\,\mathrm{GeV}\,,\ \ M_{H^\pm}\doteq91\,\mathrm{GeV}\,,
\end{equation}
which apparently contradicts the data \cite{Chatrchyan:2012vca,Aad:2012tj}. These values can be altered by changing the model parameters, the condensation scale $\Lambda$ and the number of right-handed neutrino triplets $N$.

By an order of magnitude decrease of $\Lambda$ for fixed number $N$ and $M_h=125\,\mathrm{GeV}$ we decrease the parameter $\mu_{t\nu}$ and also change the value of $\tan\beta$ according to Tab.~\ref{mutn_tanbeta_lambda}.
The values for $\Lambda$ above the Planck scale are shown only for curiosity.

By increasing the number of the right-handed neutrino triplets $N$ for a given $\Lambda$ we increase the value of $\mu_{t\nu}$ as seen in Tab.~\ref{mutnmin_mutnmax_lambda}. On the other hand the value of $\tan\beta$ is completely insensitive to the change of $N$.

Surprisingly, the number $N$ has an upper limit given by either of two conditions: the non-decoupling condition $\Lambda>M_R$ \eqref{non_decoupling_cond}, or the Higgs potential stability condition \eqref{cond_1} and \eqref{cond_2}. In the former case, increasing $N$ requires increasing $M_R$ in order to keep $m_\nu=0.2\,\mathrm{eV}$ according to the equation \eqref{nu_mass}. So for sufficiently high $N$ the mass $M_R$ runs over the condensation scale $\Lambda$. In the latter case, the increase of $N$ decreases the $\lambda_\nu$ in infrared region so that it eventually runs negative around the electroweak scale.

\begin{figure}[h]
\begin{tabular}{c}
\resizebox{0.45\textwidth}{!}{
\includegraphics{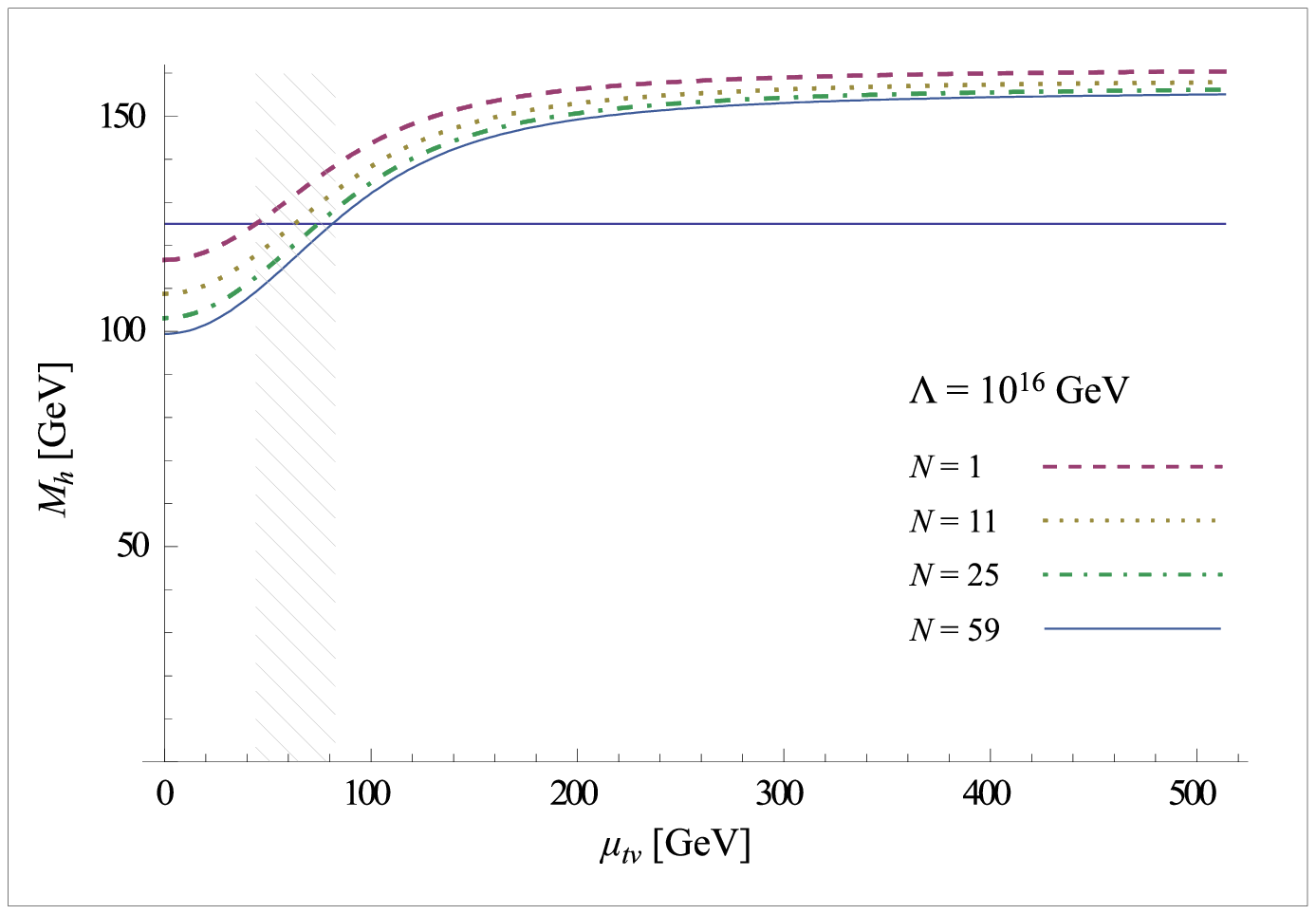}} \\
a) \\
\resizebox{0.45\textwidth}{!}{
\includegraphics{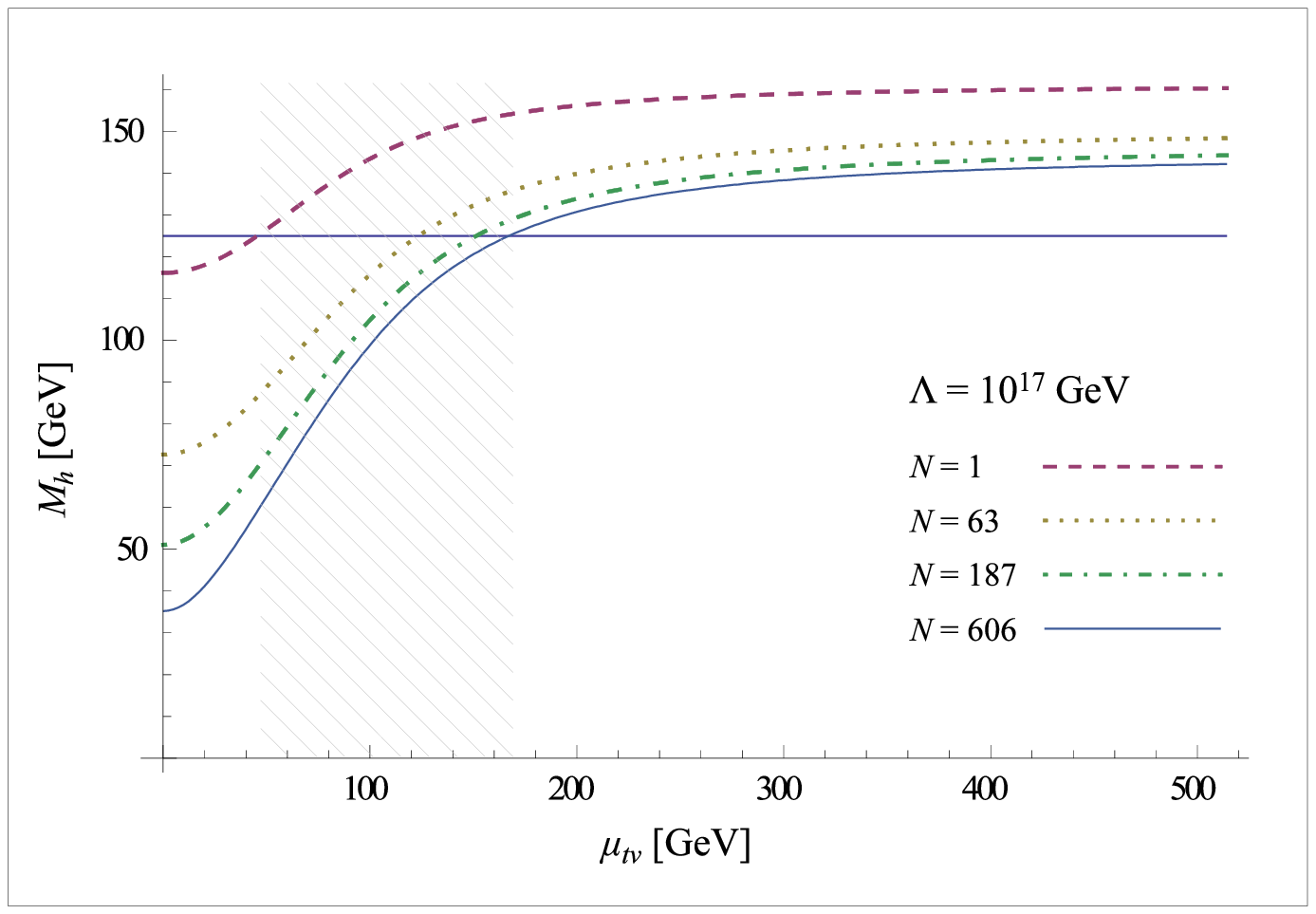}} \\
b) \\
\resizebox{0.45\textwidth}{!}{
\includegraphics{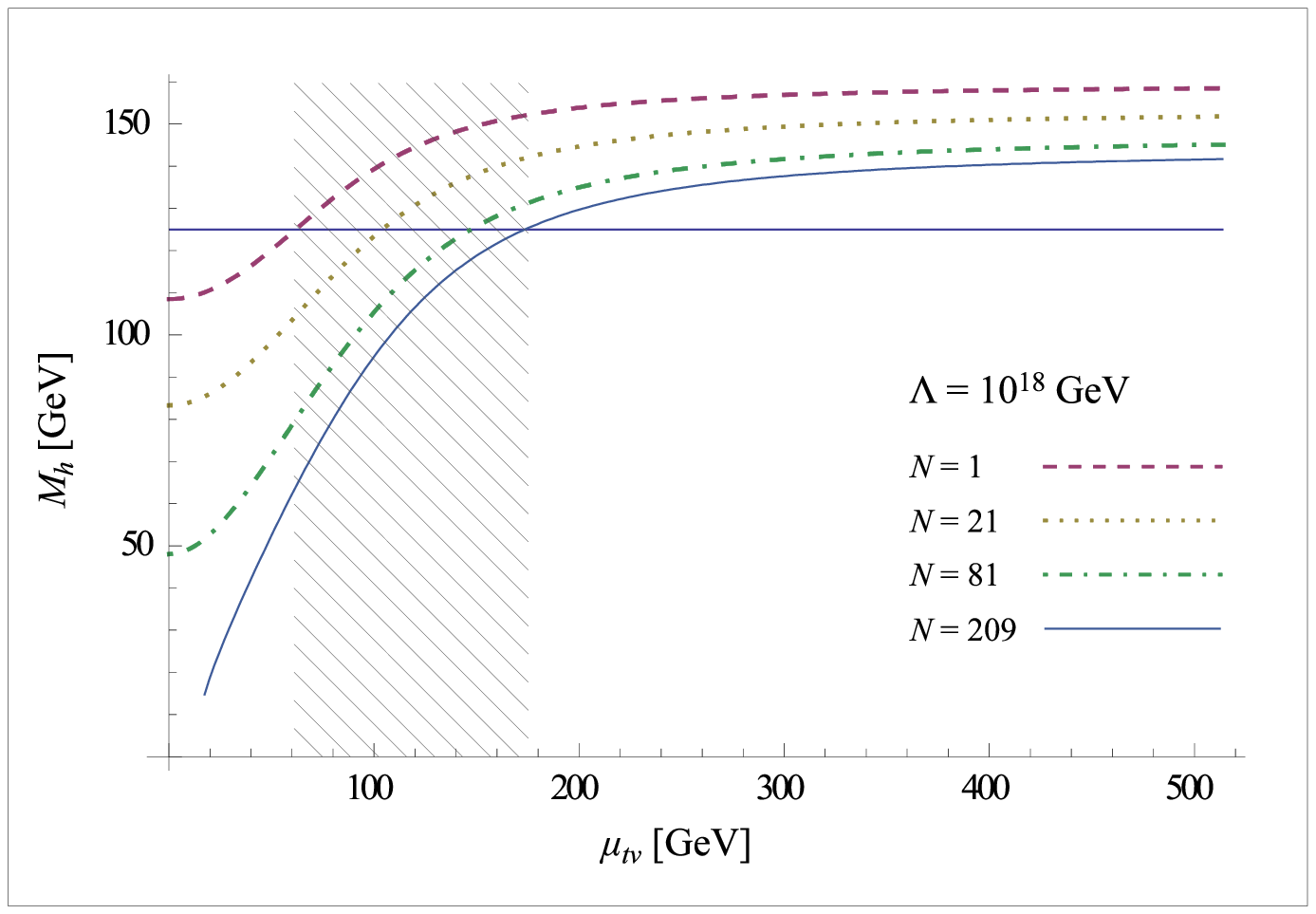}} \\
c)
\end{tabular}
\caption[]{ \small The lightest scalar Higgs mass $M_{h}$ for various numbers of right-handed neutrino triplets as a function of $\mu_{t\nu}$ for a) $\Lambda=10^{16}\,\mathrm{GeV}$, b) $\Lambda=10^{17}\,\mathrm{GeV}$, c) $\Lambda=10^{18}\,\mathrm{GeV}$, and for fixed neutrino mass to be $m_\nu=0.2\,\mathrm{eV}$. The solid horizontal line visualizes the $125\,\mathrm{GeV}$ value. The hatched area visualizes the interval of $\mu_{t\nu}$ values corresponding to $M_{h}=125\,\mathrm{GeV}$ shown in Tab.~\ref{mutnmin_mutnmax_lambda} as well. }
\label{plot_LightestHiggsMass}
\end{figure}

In Figs.~\ref{plot_LightestHiggsMass} we plot $M_{h}$ for various $N$ from $1$ to $N^\mathrm{max}$ for three cases $\Lambda=10^{16},\,10^{17},\,10^{18}\,\mathrm{GeV}$ and we read out the intervals $(\mu_{t\nu}^\mathrm{min},\mu_{t\nu}^\mathrm{max})$ of only possible values for $\mu_{t\nu}$ that correspond to $M_{h}=125\,\mathrm{GeV}$. We show it in Tab.~\ref{mutnmin_mutnmax_lambda} together with the corresponding minimal and maximal masses for $H^\pm$, $(M_{H^\pm}^\mathrm{min},M_{H^\pm}^\mathrm{max})$.

In Tab.~\ref{mutnmin_mutnmax_lambda} we show the maximum number $N^\mathrm{max}$ as well. For the case $\Lambda=10^{16}\,\mathrm{GeV}$ and $\Lambda=10^{17}\,\mathrm{GeV}$ the number $N^\mathrm{max}$ is actually not the maximal value allowed by either of conditions. It rather corresponds to maximizing the parameter $\mu_{t\nu}^\mathrm{max}$. Increasing $N$ above $N^\mathrm{max}$ causes the backward decrease of $\mu_{t\nu}$. The maximal number of right-handed neutrino triplets, above which the non-decoupling condition is broken, is $N=158$ ($N\simeq1500$) for $\Lambda=10^{16}\,\mathrm{GeV}$ ($\Lambda=10^{17}\,\mathrm{GeV}$). In the case $\Lambda=10^{18}\,\mathrm{GeV}$, the maximum number $N^\mathrm{max}=209$ is given by the Higgs potential stability.

\subsection{Light Higgs scalar coupling strengths}
\label{sec:VI.3}

\begin{figure}[t]
\begin{tabular}{c}
\resizebox{0.5\textwidth}{!}{
\includegraphics{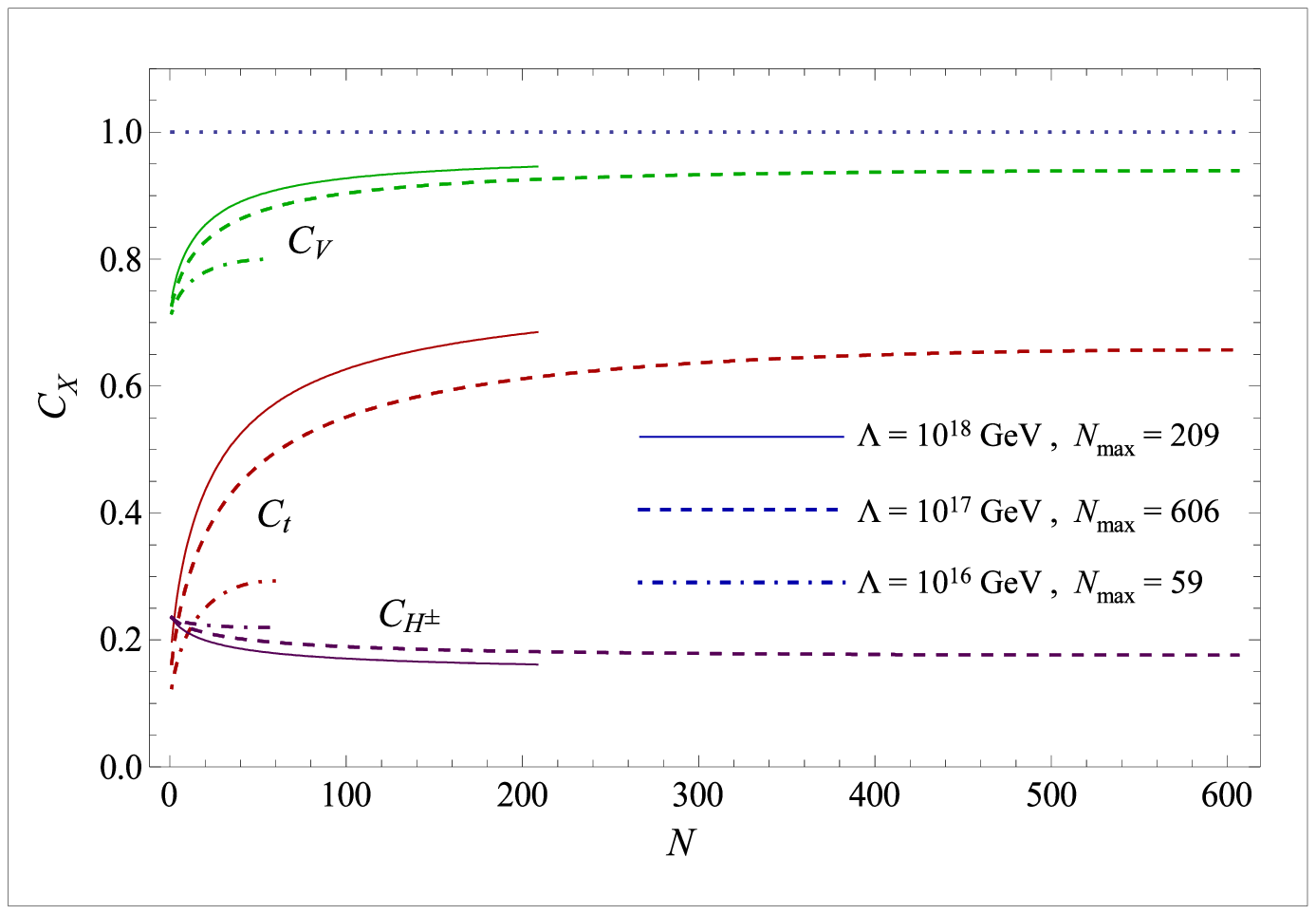}} \\
a) \\
\resizebox{0.5\textwidth}{!}{
\includegraphics{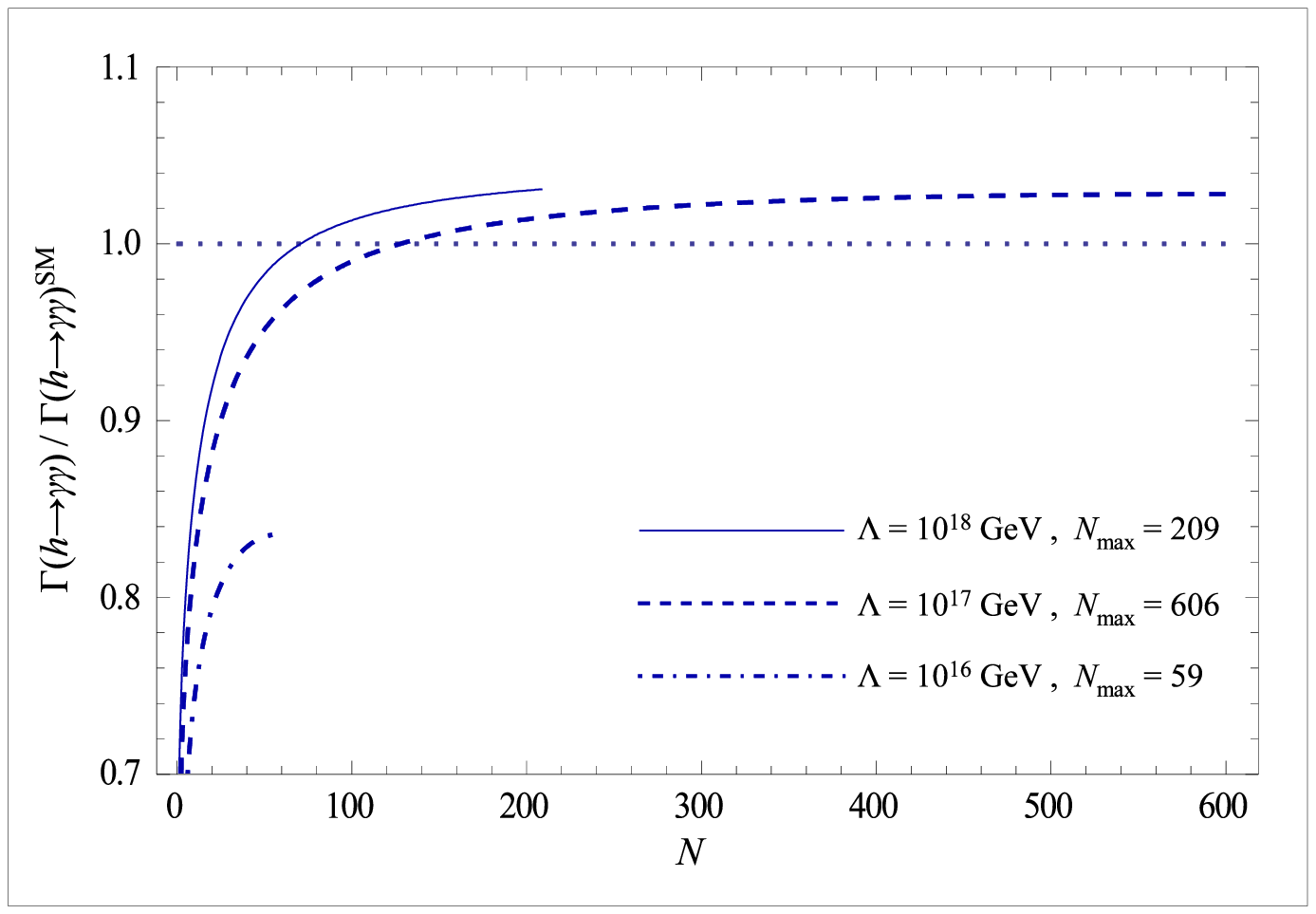}} \\
b)
\end{tabular}
\caption[]{ \small For three cases $\Lambda=10^{16},\,10^{17},\,10^{18}\,\mathrm{GeV}$ we plot a) the $N$-dependence of relative coupling parameters $C_t$, $C_V$, and $C_{H^\pm}$ defined in \eqref{coupling_parameters};
b) the dependence of the decay width $\Gamma(h\rightarrow\gamma\gamma)$ relative to the Standard model value $\Gamma(h\rightarrow\gamma\gamma)^\mathrm{SM}$ on $N$. }
\label{plot_couplings}
\end{figure}

It is mandatory to ask how strongly does the candidate for the $125\,\mathrm{GeV}$ resonance, the lighter Higgs scalar $h$, couple to the fermions and gauge bosons. We study couplings relative to the Standard model case of $h$ to $W$ and $Z$ bosons given by $C_V$, to top-quark given by $C_t$, and to charged Higgs bosons given by $C_{H^\pm}$, defined in \eqref{coupling_parameters}. We plot their dependence on the number of right-handed neutrino triplets $N$ in Fig.~\ref{plot_couplings}a) for three cases $\Lambda=10^{16},\,10^{17},\,10^{18}\,\mathrm{GeV}$. The scaling coupling factor $C_V$ approaches the Standard model value for larger $N$ in the cases $\Lambda=10^{17}\,\mathrm{GeV}$ and $\Lambda=10^{18}\,\mathrm{GeV}$. On the other hand, the coupling to the top-quark given by $C_t$ stays rather suppressed in comparison with the Standard model in all three cases.

All three coupling parameters are relevant for the loop-induced decay of $h$ to two photons. The dependence of the decay width $\Gamma(h\rightarrow\gamma\gamma)$ relative to the Standard model value $\Gamma(h\rightarrow\gamma\gamma)^\mathrm{SM}$ on the number of right-handed neutrino triplets $N$ is plotted in Fig.~\ref{plot_couplings}b). A slight enhancement occurs only for higher values of $N$ for the cases $\Lambda=10^{17}\,\mathrm{GeV}$ and $\Lambda=10^{18}\,\mathrm{GeV}$. The decay widths are calculated using the well known analytic expression to be found, e.g., in \cite{Djouadi:2005gj}.

\begin{table}[t]
\begin{center}
\begin{tabular}{c|cc}
\hline\noalign{\smallskip}
$\Lambda\,[\mathrm{GeV}]$ & $\mu_{t\nu}\,[\mathrm{GeV}]$ & $\tan{\beta}$  \\
\noalign{\smallskip}\hline\noalign{\smallskip}
$10^{16}$ & $44$ & $1.183$  \\
$10^{17}$ & $54$ & $1.215$  \\
$10^{18}$ & $62$ & $1.245$  \\
\noalign{\smallskip}\hline\noalign{\smallskip}
$10^{24}$ & $86$ & $1.401$  \\
$10^{40}$ & $101$ & $1.624$ \\
\noalign{\smallskip}\hline
\end{tabular}
\end{center}
\caption{\small Values of $\mu_{t\nu}$ and $\tan{\beta}$ depending on $\Lambda$ while keeping $m_\nu=0.2\,\mathrm{eV}$ and $M_{h}=125\,\mathrm{GeV}$ for $N=1$. }
\label{mutn_tanbeta_lambda}
\end{table}

\begin{table}[t]
\begin{center}
\begin{tabular}{rc|cc|cc}
\hline\noalign{\smallskip}
 &  & $\mu_{t\nu}^\mathrm{min}$  & $\mu_{t\nu}^\mathrm{max}$ & $M_{H^\pm}^\mathrm{min}$ & $M_{H^\pm}^\mathrm{max}$  \\
$\Lambda\,[\mathrm{GeV}]$ \vline& $N^\mathrm{max}$ & $[\mathrm{GeV}]$  & $[\mathrm{GeV}]$ & $[\mathrm{GeV}]$ & $[\mathrm{GeV}]$  \\
\noalign{\smallskip}\hline\noalign{\smallskip}
$10^{16}$\ \ \vline& $59$  & $44$ & $81$  & $66$ & $117$ \\
$10^{17}$\ \ \vline& $606$ & $54$ & $164$ & $80$ & $234$ \\
$10^{18}$\ \ \vline& $209$ & $62$ & $173$ & $92$ & $249$ \\
\noalign{\smallskip}\hline
\end{tabular}
\end{center}
\caption{\small We show the maximum number of right-handed neutrino triplets $N^\mathrm{max}$ for three cases $\Lambda=10^{16},\,10^{17},\,10^{18}\,\mathrm{GeV}$. Next, we show the only intervals for $\mu_{t\nu}$,  $(\mu_{t\nu}^\mathrm{min},\mu_{t\nu}^\mathrm{max})$, allowed by $M_{h}=125\,\mathrm{GeV}$. Finally, we show the corresponding intervals for charged Higgs boson masses $(M_{H^\pm}^\mathrm{min},M_{H^\pm}^\mathrm{max})$. }
\label{mutnmin_mutnmax_lambda}
\end{table}

\section{Discussion}\label{SecVII}

In this work we have chosen the two-Higgs-doublet model as the next-to-simplest model accommodating the leading idea of our interest -- the top-quark and neutrino condensation as a sufficient source of the electroweak symmetry breaking -- as it was proposed in \cite{Antusch:2002xh}.

Our analysis shows that it is possible to simultaneously reproduce correct values for the top-quark mass $m_t$, the electroweak scale $v$, the $125\,\mathrm{GeV}$ boson mass, and the neutrino mass $m_\nu$ below observational upper limit, despite rather limited manoeuvring space for participating parameters.

The number of right-handed neutrino types participating on the seesaw mechanism is not constrained phenomenologically by any upper limit \cite{Ellis:2007wz,Heeck:2012fw}. The model however exhibits an interesting feature providing the upper limit on that number.

Next, we present two aspects of the model which are relevant for a present phenomenology: the mass spectrum of the additional Higgs bosons $H$, $A$ and $H^\pm$, and the coupling strengths of the $125\,\mathrm{GeV}$ Higgs boson to the top-quark and gauge bosons. We study their dependence on $N$ and on the condensation scale $\Lambda$. Generally speaking, the higher values of $N$ and $\Lambda$ are preferred, because they lead to higher values of additional Higgs boson masses, and to the coupling strengths closer to the Standard model values. For example, for $\Lambda=10^{18}$ and $N=100$ we obtain the charged Higgs boson mass $M_{H^\pm}\doteq223\,\mathrm{GeV}$ and the coupling constant of $h$ to $W$ and $Z$ at $93\%$ level of the Standard model value.

The confrontation of these two aspects with the experimental constraints in the following two subsections however should be taken with a grain of salt for three reasons. First, the model analyzed in this work is only a semi-realistic model: it ignores the mass generation of fermions other than the top-quark and neutrinos. Second, it is subject to simplification of the neutrino sector. Third, it is not possible to directly link our model to one of the standard types of two-Higgs-doublet models, for which the data analyses are available. They usually deal with full arsenal of Yukawa interactions with charged fermions. In our model, we avoid Yukawa interactions with lighter fermions in favour of the Yukawa coupling of one of the Higgs doublets to right-handed neutrinos.

\subsection{Mass of charged Higgs boson}
\label{sec:VII.1}

The values of $M_{H^\pm}$ and $\tan\beta$ accessible in the model for the higher number of right-handed neutrino triplets $N$ lie in the ranges
\begin{equation}
M_{H^\pm}\simeq(200-250)\mathrm{GeV}\,,\ \ \tan\beta\simeq(1.2-1.25)\,,
\end{equation}
see Tab.~\ref{mutn_tanbeta_lambda} and Tab.~\ref{mutnmin_mutnmax_lambda}.

Direct searches for the charged Higgs boson at LHC give a lower limit for its mass $M_{H^\pm}>160\,\mathrm{GeV}$ \cite{Chatrchyan:2012vca,Aad:2012tj}. The excluded mass interval corresponds to the below-threshold production $t\rightarrow H^+b$. The analyses are made under assumption of $100\%$ decay ratio via $H^+\rightarrow\tau\nu$ and within one of the MSSM scenario. The lower limit for $M_{H^\pm}$ translates in our model into the lower limit for $N$
\begin{eqnarray}
N>38  \ \ & \mathrm{for} & \ \ \Lambda=10^{17}\,\mathrm{GeV} \,, \\
N>20  \ \ & \mathrm{for} & \ \ \Lambda=10^{18}\,\mathrm{GeV} \,,
\end{eqnarray}
while the case $\Lambda=10^{16}\,\mathrm{GeV}$ is excluded.

Indirect searches in $B$-physics are more stringent in setting the lower limits, but they are more model-dependent at the same time. For review see \cite{Branco:2011iw}. For example, the limit $M_{H^\pm}>300-400\,\mathrm{GeV}$ from the $B\rightarrow X_s\gamma$ decay is set for type-II two-Higgs-doublet model, and the limit $M_{H^\pm}>160\,\mathrm{GeV}$ ($M_{H^\pm}>500\,\mathrm{GeV}$)\footnote{The limit in parenthesis follows from $B\rightarrow X_s\gamma$ but it is very sensitive to assumptions and to input parameters.} from $Z\rightarrow b\bar{b}$ and from $B$- and $K$-meson mixing is set for type-I two-Higgs-doublet model. Even though it is rather speculative to apply those constraints to our model, they indicate that the model appears to be at the edge of viability.

\subsection{Production and decays of lighter Higgs scalar}
\label{sec:VII.2}

In order to successfully identify the lighter Higgs scalar $h$ with the observed $125\,\mathrm{GeV}$ boson, it should exhibit coupling properties to other particles which leads to the observed phenomena.

Provided the higher number of right-handed neutrino triplets $N>100$ and $\Lambda=10^{18}\,\mathrm{GeV}$, the $h$ scaling coupling factors \eqref{coupling_parameters} characteristic for the model are on the level
\begin{equation}\label{CtCV}
C_t\simeq(0.63-0.69) \,,\ \ C_V\simeq(0.93-0.95) \,,\ \
\end{equation}
see Fig.~\ref{plot_couplings}a).

This result is to be compared with the ATLAS \cite{ATLAS:2012wma} and CMS \cite{CMS:2012wwa} results for corresponding quantities.\footnote{In the ATLAS and CMS analyses the overall fermion scaling factor $C_F$ is used, instead of $C_t$, which scales only the top-quark Yukawa interaction in our model. In any case $C_t=C_F$.} For example, the gluon fusion cross-section $gg\rightarrow h$ being induced by top-quark loop scales with the factor $C_{t}^2$, or the partial decay width for $h\rightarrow WW$ scales with the factor $C_{V}^2$. The best-fit values over all observed production-decay modes are
\begin{eqnarray}
(C_t,C_V) & \simeq & (1.0,1.2) \ \ \ \mathrm{ATLAS} \,, \\
(C_t,C_V) & \simeq & (0.5,1.0) \ \ \ \mathrm{CMS} \,.
\end{eqnarray}
The point \eqref{CtCV} lies within the ATLAS $95\,\%$ confidence level range and within the CMS less than $68\,\%$ confidence level range. The analyses where made under assumption that no non-Standard model particles contribute to the total decay width, what is reasonable assumption for our model, as they all are heavier than the lighter Higgs scalar $h$.

Out of the individual decay channels, we discuss $h\rightarrow\gamma\gamma$. Provided the higher number of right-handed neutrino triplets $N>100$ and $\Lambda=10^{18}\,\mathrm{GeV}$, the enhancement of the partial decay width $\Gamma(h\rightarrow\gamma\gamma)$ with respect to the Standard model can be achieved at the level of $(1-3)\%$, see Fig.~\ref{plot_couplings}b),
\begin{equation}
\frac{\Gamma(h\rightarrow\gamma\gamma)}{\Gamma(h\rightarrow\gamma\gamma)^\mathrm{SM}}\simeq(1.01-1.03) \,.
\end{equation}
The signal strength $\mu$ for $h\rightarrow\gamma\gamma$ channel is measured with a $2\sigma$ excess with respect to the Standard model \cite{ATLAS:2012wma,CMS:2012wwa}
\begin{equation}
\mu(h\rightarrow\gamma\gamma) = \frac{\sigma_h}{\sigma_{h}^\mathrm{SM}}\frac{\Gamma(h\rightarrow\gamma\gamma)/\Gamma_\mathrm{tot.}}{\Gamma(h\rightarrow\gamma\gamma)^\mathrm{SM}/\Gamma_{\mathrm{tot.}}^\mathrm{SM}}\sim(1.5-2.0)\,.
\end{equation}
The main part of the $h$ production cross-section $\sigma_h$ is given by gluon fusion cross-section $\sigma(gg\rightarrow h)$ which scales as $C_{t}^2\simeq(0.39-0.47)$ in our model. This suppression can be however compensated by the suppression of the total decay width $\Gamma_\mathrm{tot.}$, which scales as a linear combination of the $C_{t}^2$ and $C_{V}^2$ factors, both presenting a suppression. According to its parameter setting the model does not profit from the presence of the charged Higgs boson in order to enhance significantly the $h\rightarrow\gamma\gamma$ signal strength with respect to the Standard model prediction.

\section{Conclusions}\label{SecVIII}

We have analyzed the top-quark and neutrino condensation scenario for the electroweak symmetry breaking within a reasonably simplified semi-realistic model with two composite Higgs doublets. We have demonstrated that it is possible to reproduce a mass spectrum of top-quark, neutrinos and observed bosons.

The lighter Higgs scalar is identified with $125\,\mathrm{GeV}$ particle. However some indirect constraints on the masses of yet unobserved additional Higgs bosons indicate that the model appears to be at the edge of viability. The coupling strengths of $h$ differs from the Standard model values, but the experimental data have not a decisive power yet in this respect. If we keep only well established and model independent constraints then there remains some room for the top-quark and neutrino condensation scenario, provided that the condensation scale is $\Lambda\sim10^{17-18}\,\mathrm{GeV}$ and the number of right-handed neutrinos participating on the seesaw mechanism is ${\cal O}(100-1000)$.

There are two detail-independent predictions of the scenario. First, the $h$ has rather big admixture of the neutrinos given by $\alpha\sim-0.8$. The mixing factor suppresses its Yukawa coupling with the top-quark and eventually with other charged fermions at the level of $\sim60\,\%$ in comparison with the Standard model. Second, the scenario provides an upper limit on the additional Higgs bosons which is rather low $<250\,\mathrm{GeV}$. Through both predictions the scenario should be easily and definitely falsifiable by delivering more data from LHC in the near future.

\begin{acknowledgments}
The author gratefully acknowledges discussions with J. Ho\v{s}ek and P. Bene\v{s}. The author is also grateful for meeting with F. Sannino and M. Lindner and discussing the issue with them. The work was supported by the Research Program MSM6840770029, by the MEIS of Czech Republic LM2011027 and by the project of International Cooperation ATLAS-CERN LA08032. The work was also supported by TJ Balvan Praha.
\end{acknowledgments}

\bibliography{references}

\begin{thebibliography}{39}
\expandafter\ifx\csname natexlab\endcsname\relax\def\natexlab#1{#1}\fi
\expandafter\ifx\csname bibnamefont\endcsname\relax
  \def\bibnamefont#1{#1}\fi
\expandafter\ifx\csname bibfnamefont\endcsname\relax
  \def\bibfnamefont#1{#1}\fi
\expandafter\ifx\csname citenamefont\endcsname\relax
  \def\citenamefont#1{#1}\fi
\expandafter\ifx\csname url\endcsname\relax
  \def\url#1{\texttt{#1}}\fi
\expandafter\ifx\csname urlprefix\endcsname\relax\def\urlprefix{URL }\fi
\providecommand{\bibinfo}[2]{#2}
\providecommand{\eprint}[2][]{\url{#2}}

\bibitem[{\citenamefont{Aad et~al.}(2012{\natexlab{a}})}]{125Higgs:2012gk}
\bibinfo{author}{\bibfnamefont{G.}~\bibnamefont{Aad}} \bibnamefont{et~al.}
  (\bibinfo{collaboration}{ATLAS}), \bibinfo{journal}{Phys. Lett.}
  \textbf{\bibinfo{volume}{B716}}, \bibinfo{pages}{1}
  (\bibinfo{year}{2012}{\natexlab{a}}), \eprint{1207.7214}.

\bibitem[{\citenamefont{Chatrchyan
  et~al.}(2012{\natexlab{a}})}]{125Higgs:2012gu}
\bibinfo{author}{\bibfnamefont{S.}~\bibnamefont{Chatrchyan}}
  \bibnamefont{et~al.} (\bibinfo{collaboration}{CMS}),
  \bibinfo{journal}{Phys.Lett.B} \textbf{\bibinfo{volume}{B716}},
  \bibinfo{pages}{30} (\bibinfo{year}{2012}{\natexlab{a}}), \eprint{1207.7235}.

\bibitem[{\citenamefont{Degrassi et~al.}(2012)\citenamefont{Degrassi, Di~Vita,
  Elias-Miro, Espinosa, Giudice et~al.}}]{Degrassi:2012ry}
\bibinfo{author}{\bibfnamefont{G.}~\bibnamefont{Degrassi}},
  \bibinfo{author}{\bibfnamefont{S.}~\bibnamefont{Di~Vita}},
  \bibinfo{author}{\bibfnamefont{J.}~\bibnamefont{Elias-Miro}},
  \bibinfo{author}{\bibfnamefont{J.~R.} \bibnamefont{Espinosa}},
  \bibinfo{author}{\bibfnamefont{G.~F.} \bibnamefont{Giudice}},
  \bibnamefont{et~al.}, \bibinfo{journal}{JHEP}
  \textbf{\bibinfo{volume}{1208}}, \bibinfo{pages}{098} (\bibinfo{year}{2012}),
  \eprint{1205.6497}.

\bibitem[{\citenamefont{Bednyakov et~al.}(2013)\citenamefont{Bednyakov,
  Pikelner, and Velizhanin}}]{Bednyakov:2013eba}
\bibinfo{author}{\bibfnamefont{A.}~\bibnamefont{Bednyakov}},
  \bibinfo{author}{\bibfnamefont{A.}~\bibnamefont{Pikelner}}, \bibnamefont{and}
  \bibinfo{author}{\bibfnamefont{V.}~\bibnamefont{Velizhanin}}
  (\bibinfo{year}{2013}), \eprint{1303.4364}.

\bibitem[{\citenamefont{Chetyrkin and Zoller}(2013)}]{Chetyrkin:2013wya}
\bibinfo{author}{\bibfnamefont{K.}~\bibnamefont{Chetyrkin}} \bibnamefont{and}
  \bibinfo{author}{\bibfnamefont{M.}~\bibnamefont{Zoller}},
  \bibinfo{journal}{JHEP} \textbf{\bibinfo{volume}{1304}}, \bibinfo{pages}{091}
  (\bibinfo{year}{2013}), \eprint{1303.2890}.

\bibitem[{\citenamefont{Ho\v{s}ek}(1985)}]{Hosek:1985jr}
\bibinfo{author}{\bibfnamefont{J.}~\bibnamefont{Ho\v{s}ek}},
  \bibinfo{journal}{CERN-TH-4104/85}  (\bibinfo{year}{1985}).

\bibitem[{\citenamefont{Miransky et~al.}(1989)\citenamefont{Miransky,
  Tanabashi, and Yamawaki}}]{Miransky:1988xi}
\bibinfo{author}{\bibfnamefont{V.~A.} \bibnamefont{Miransky}},
  \bibinfo{author}{\bibfnamefont{M.}~\bibnamefont{Tanabashi}},
  \bibnamefont{and} \bibinfo{author}{\bibfnamefont{K.}~\bibnamefont{Yamawaki}},
  \bibinfo{journal}{Phys. Lett.} \textbf{\bibinfo{volume}{B221}},
  \bibinfo{pages}{177} (\bibinfo{year}{1989}).

\bibitem[{\citenamefont{Bardeen et~al.}(1990)\citenamefont{Bardeen, Hill, and
  Lindner}}]{Bardeen:1989ds}
\bibinfo{author}{\bibfnamefont{W.~A.} \bibnamefont{Bardeen}},
  \bibinfo{author}{\bibfnamefont{C.~T.} \bibnamefont{Hill}}, \bibnamefont{and}
  \bibinfo{author}{\bibfnamefont{M.}~\bibnamefont{Lindner}},
  \bibinfo{journal}{Phys. Rev.} \textbf{\bibinfo{volume}{D41}},
  \bibinfo{pages}{1647} (\bibinfo{year}{1990}).

\bibitem[{\citenamefont{Martin}(1991)}]{Martin:1991xw}
\bibinfo{author}{\bibfnamefont{S.~P.} \bibnamefont{Martin}},
  \bibinfo{journal}{Phys. Rev.} \textbf{\bibinfo{volume}{D44}},
  \bibinfo{pages}{2892} (\bibinfo{year}{1991}).

\bibitem[{\citenamefont{Cvetic and Kim}(1994{\natexlab{a}})}]{Cvetic:1992ps}
\bibinfo{author}{\bibfnamefont{G.}~\bibnamefont{Cvetic}} \bibnamefont{and}
  \bibinfo{author}{\bibfnamefont{C.}~\bibnamefont{Kim}}, \bibinfo{journal}{Mod.
  Phys. Lett.} \textbf{\bibinfo{volume}{A9}}, \bibinfo{pages}{289}
  (\bibinfo{year}{1994}{\natexlab{a}}).

\bibitem[{\citenamefont{Antusch et~al.}(2003)\citenamefont{Antusch, Kersten,
  Lindner, and Ratz}}]{Antusch:2002xh}
\bibinfo{author}{\bibfnamefont{S.}~\bibnamefont{Antusch}},
  \bibinfo{author}{\bibfnamefont{J.}~\bibnamefont{Kersten}},
  \bibinfo{author}{\bibfnamefont{M.}~\bibnamefont{Lindner}}, \bibnamefont{and}
  \bibinfo{author}{\bibfnamefont{M.}~\bibnamefont{Ratz}},
  \bibinfo{journal}{Nucl. Phys.} \textbf{\bibinfo{volume}{B658}},
  \bibinfo{pages}{203} (\bibinfo{year}{2003}), \eprint{hep-ph/0211385}.

\bibitem[{\citenamefont{Ho\v{s}ek}(1982)}]{Hosek:1982cz}
\bibinfo{author}{\bibfnamefont{J.}~\bibnamefont{Ho\v{s}ek}},
  \bibinfo{journal}{Submitted to Nucl. Phys.}  (\bibinfo{year}{1982}),
  \bibinfo{note}{preprint JINR-E2-82-542, submitted to 21st Int. Conf. on High
  Energy Physics, Paris, France, Jul 26-31, 1982}.

\bibitem[{\citenamefont{Kimura and Munakata}(1984)}]{Kimura:1984zv}
\bibinfo{author}{\bibfnamefont{K.}~\bibnamefont{Kimura}} \bibnamefont{and}
  \bibinfo{author}{\bibfnamefont{H.}~\bibnamefont{Munakata}},
  \bibinfo{journal}{KOBE-84-04}  (\bibinfo{year}{1984}).

\bibitem[{\citenamefont{Nagoshi et~al.}(1991)\citenamefont{Nagoshi, Nakanishi,
  and Tanaka}}]{Nagoshi:1990wk}
\bibinfo{author}{\bibfnamefont{Y.}~\bibnamefont{Nagoshi}},
  \bibinfo{author}{\bibfnamefont{K.}~\bibnamefont{Nakanishi}},
  \bibnamefont{and} \bibinfo{author}{\bibfnamefont{S.}~\bibnamefont{Tanaka}},
  \bibinfo{journal}{Prog. Theor. Phys.} \textbf{\bibinfo{volume}{85}},
  \bibinfo{pages}{131} (\bibinfo{year}{1991}).

\bibitem[{\citenamefont{Cvetic and Kim}(1994{\natexlab{b}})}]{Cvetic:1992xn}
\bibinfo{author}{\bibfnamefont{G.}~\bibnamefont{Cvetic}} \bibnamefont{and}
  \bibinfo{author}{\bibfnamefont{C.}~\bibnamefont{Kim}}, \bibinfo{journal}{Int.
  J. Mod. Phys.} \textbf{\bibinfo{volume}{A9}}, \bibinfo{pages}{1495}
  (\bibinfo{year}{1994}{\natexlab{b}}).

\bibitem[{\citenamefont{Gribov}(1994)}]{Gribov:1994jy}
\bibinfo{author}{\bibfnamefont{V.}~\bibnamefont{Gribov}},
  \bibinfo{journal}{Phys. Lett.} \textbf{\bibinfo{volume}{B336}},
  \bibinfo{pages}{243} (\bibinfo{year}{1994}), \eprint{hep-ph/9407269}.

\bibitem[{\citenamefont{Bashford}(2003)}]{Bashford:2003yg}
\bibinfo{author}{\bibfnamefont{J.~D.} \bibnamefont{Bashford}}
  (\bibinfo{year}{2003}), \eprint{hep-ph/0310309}.

\bibitem[{\citenamefont{Brauner and Ho\v{s}ek}(2004)}]{Brauner:2004kg}
\bibinfo{author}{\bibfnamefont{T.}~\bibnamefont{Brauner}} \bibnamefont{and}
  \bibinfo{author}{\bibfnamefont{J.}~\bibnamefont{Ho\v{s}ek}}
  (\bibinfo{year}{2004}), \eprint{hep-ph/0407339}.

\bibitem[{\citenamefont{Bene\v{s} et~al.}(2009)\citenamefont{Bene\v{s},
  Brauner, and Smetana}}]{Benes:2008ir}
\bibinfo{author}{\bibfnamefont{P.}~\bibnamefont{Bene\v{s}}},
  \bibinfo{author}{\bibfnamefont{T.}~\bibnamefont{Brauner}}, \bibnamefont{and}
  \bibinfo{author}{\bibfnamefont{A.}~\bibnamefont{Smetana}},
  \bibinfo{journal}{J. Phys.} \textbf{\bibinfo{volume}{G36}},
  \bibinfo{pages}{115004} (\bibinfo{year}{2009}), \eprint{0806.2565}.

\bibitem[{\citenamefont{Wetterich}(2006)}]{Wetterich:2006ii}
\bibinfo{author}{\bibfnamefont{C.}~\bibnamefont{Wetterich}},
  \bibinfo{journal}{Phys. Rev.} \textbf{\bibinfo{volume}{D74}},
  \bibinfo{pages}{095009} (\bibinfo{year}{2006}), \eprint{hep-ph/0607051}.

\bibitem[{\citenamefont{Schwindt and Wetterich}(2010)}]{Schwindt:2008gj}
\bibinfo{author}{\bibfnamefont{J.-M.} \bibnamefont{Schwindt}} \bibnamefont{and}
  \bibinfo{author}{\bibfnamefont{C.}~\bibnamefont{Wetterich}},
  \bibinfo{journal}{Phys. Rev.} \textbf{\bibinfo{volume}{D81}},
  \bibinfo{pages}{055005} (\bibinfo{year}{2010}), \eprint{0812.4223}.

\bibitem[{\citenamefont{Ho\v{s}ek}(2009)}]{Hosek:2009ys}
\bibinfo{author}{\bibfnamefont{J.}~\bibnamefont{Ho\v{s}ek}}
  (\bibinfo{year}{2009}), \eprint{0909.0629}.

\bibitem[{\citenamefont{Ho\v{s}ek}(2011)}]{Hosek:NagoyaProceeding}
\bibinfo{author}{\bibfnamefont{J.}~\bibnamefont{Ho\v{s}ek}}, in
  \emph{\bibinfo{booktitle}{Proceedings of the Workshop in Honor of Toshihide
  Maskawa's 70th Birthday and 35th Anniversary of Dynamical Symmetry Breaking
  in SCGT}}, edited by
  \bibinfo{editor}{\bibfnamefont{H.}~\bibnamefont{Fukaya}},
  \bibinfo{editor}{\bibfnamefont{M.}~\bibnamefont{Harada}},
  \bibinfo{editor}{\bibfnamefont{M.}~\bibnamefont{Tanabashi}},
  \bibnamefont{and} \bibinfo{editor}{\bibfnamefont{K.}~\bibnamefont{Yamawaki}}
  (\bibinfo{publisher}{World Scientific}, \bibinfo{year}{2011}), pp.
  \bibinfo{pages}{191--197}.

\bibitem[{\citenamefont{Bene\v{s} et~al.}(2011)\citenamefont{Bene\v{s},
  Ho\v{s}ek, and Smetana}}]{Benes:2011gi}
\bibinfo{author}{\bibfnamefont{P.}~\bibnamefont{Bene\v{s}}},
  \bibinfo{author}{\bibfnamefont{J.}~\bibnamefont{Ho\v{s}ek}},
  \bibnamefont{and} \bibinfo{author}{\bibfnamefont{A.}~\bibnamefont{Smetana}}
  (\bibinfo{year}{2011}), \eprint{1101.3456}.

\bibitem[{\citenamefont{Smetana}(2013)}]{Smetana:2011tj}
\bibinfo{author}{\bibfnamefont{A.}~\bibnamefont{Smetana}},
  \bibinfo{journal}{JHEP} \textbf{\bibinfo{volume}{1304}}, \bibinfo{pages}{139}
  (\bibinfo{year}{2013}), \eprint{1104.1935}.

\bibitem[{\citenamefont{Ellis and Lebedev}(2007)}]{Ellis:2007wz}
\bibinfo{author}{\bibfnamefont{J.~R.} \bibnamefont{Ellis}} \bibnamefont{and}
  \bibinfo{author}{\bibfnamefont{O.}~\bibnamefont{Lebedev}},
  \bibinfo{journal}{Phys. Lett.} \textbf{\bibinfo{volume}{B653}},
  \bibinfo{pages}{411} (\bibinfo{year}{2007}), \eprint{0707.3419}.

\bibitem[{\citenamefont{Heeck}(2012)}]{Heeck:2012fw}
\bibinfo{author}{\bibfnamefont{J.}~\bibnamefont{Heeck}},
  \bibinfo{journal}{Phys. Rev.} \textbf{\bibinfo{volume}{D86}},
  \bibinfo{pages}{093023} (\bibinfo{year}{2012}), \eprint{1207.5521}.

\bibitem[{\citenamefont{Eisele}(2008)}]{Eisele:2007ws}
\bibinfo{author}{\bibfnamefont{M.-T.} \bibnamefont{Eisele}},
  \bibinfo{journal}{Phys. Rev.} \textbf{\bibinfo{volume}{D77}},
  \bibinfo{pages}{043510} (\bibinfo{year}{2008}), \eprint{0706.0200}.

\bibitem[{\citenamefont{Feldstein and Klemm}(2012)}]{Feldstein:2011ck}
\bibinfo{author}{\bibfnamefont{B.}~\bibnamefont{Feldstein}} \bibnamefont{and}
  \bibinfo{author}{\bibfnamefont{W.}~\bibnamefont{Klemm}},
  \bibinfo{journal}{Phys. Rev.} \textbf{\bibinfo{volume}{D85}},
  \bibinfo{pages}{053007} (\bibinfo{year}{2012}), \eprint{1111.6690}.

\bibitem[{\citenamefont{Stratonovich}(1957)}]{Stratonovich:1957aa}
\bibinfo{author}{\bibfnamefont{R.~L.} \bibnamefont{Stratonovich}},
  \bibinfo{journal}{Sov. Phys. Doklady} \textbf{\bibinfo{volume}{2}},
  \bibinfo{pages}{416} (\bibinfo{year}{1957}).

\bibitem[{\citenamefont{Hubbard}(1959)}]{Hubbard:1959aa}
\bibinfo{author}{\bibfnamefont{J.}~\bibnamefont{Hubbard}},
  \bibinfo{journal}{{Phys. Rev. Lett.}} \textbf{\bibinfo{volume}{3}},
  \bibinfo{pages}{77} (\bibinfo{year}{1959}).

\bibitem[{\citenamefont{Luty}(1990)}]{Luty:1990bg}
\bibinfo{author}{\bibfnamefont{M.~A.} \bibnamefont{Luty}},
  \bibinfo{journal}{Phys. Rev.} \textbf{\bibinfo{volume}{D41}},
  \bibinfo{pages}{2893} (\bibinfo{year}{1990}).

\bibitem[{\citenamefont{Hill et~al.}(1985)\citenamefont{Hill, Leung, and
  Rao}}]{Hill:1985tg}
\bibinfo{author}{\bibfnamefont{C.~T.} \bibnamefont{Hill}},
  \bibinfo{author}{\bibfnamefont{C.~N.} \bibnamefont{Leung}}, \bibnamefont{and}
  \bibinfo{author}{\bibfnamefont{S.}~\bibnamefont{Rao}},
  \bibinfo{journal}{Nucl. Phys.} \textbf{\bibinfo{volume}{B262}},
  \bibinfo{pages}{517} (\bibinfo{year}{1985}).

\bibitem[{\citenamefont{Chatrchyan
  et~al.}(2012{\natexlab{b}})}]{Chatrchyan:2012vca}
\bibinfo{author}{\bibfnamefont{S.}~\bibnamefont{Chatrchyan}}
  \bibnamefont{et~al.} (\bibinfo{collaboration}{CMS}), \bibinfo{journal}{JHEP}
  \textbf{\bibinfo{volume}{1207}}, \bibinfo{pages}{143}
  (\bibinfo{year}{2012}{\natexlab{b}}), \eprint{1205.5736}.

\bibitem[{\citenamefont{Aad et~al.}(2012{\natexlab{b}})}]{Aad:2012tj}
\bibinfo{author}{\bibfnamefont{G.}~\bibnamefont{Aad}} \bibnamefont{et~al.}
  (\bibinfo{collaboration}{ATLAS}), \bibinfo{journal}{JHEP}
  \textbf{\bibinfo{volume}{1206}}, \bibinfo{pages}{039}
  (\bibinfo{year}{2012}{\natexlab{b}}), \eprint{1204.2760}.

\bibitem[{\citenamefont{Djouadi}(2008)}]{Djouadi:2005gj}
\bibinfo{author}{\bibfnamefont{A.}~\bibnamefont{Djouadi}},
  \bibinfo{journal}{Phys. Rept.} \textbf{\bibinfo{volume}{459}},
  \bibinfo{pages}{1} (\bibinfo{year}{2008}), \eprint{hep-ph/0503173}.

\bibitem[{\citenamefont{Branco et~al.}(2012)\citenamefont{Branco, Ferreira,
  Lavoura, Rebelo, Sher et~al.}}]{Branco:2011iw}
\bibinfo{author}{\bibfnamefont{G.}~\bibnamefont{Branco}},
  \bibinfo{author}{\bibfnamefont{P.}~\bibnamefont{Ferreira}},
  \bibinfo{author}{\bibfnamefont{L.}~\bibnamefont{Lavoura}},
  \bibinfo{author}{\bibfnamefont{M.}~\bibnamefont{Rebelo}},
  \bibinfo{author}{\bibfnamefont{M.}~\bibnamefont{Sher}}, \bibnamefont{et~al.},
  \bibinfo{journal}{Phys. Rept.} \textbf{\bibinfo{volume}{516}},
  \bibinfo{pages}{1} (\bibinfo{year}{2012}), \eprint{1106.0034}.

\bibitem[{\citenamefont{Aad et~al.}(2012{\natexlab{c}})}]{ATLAS:2012wma}
\bibinfo{author}{\bibfnamefont{G.}~\bibnamefont{Aad}} \bibnamefont{et~al.}
  (\bibinfo{collaboration}{ATLAS}), \bibinfo{journal}{ATLAS-CONF-2012-127,
  ATLAS-COM-CONF-2012-161}  (\bibinfo{year}{2012}{\natexlab{c}}).

\bibitem[{\citenamefont{Chatrchyan et~al.}(2012{\natexlab{c}})}]{CMS:2012wwa}
\bibinfo{author}{\bibfnamefont{S.}~\bibnamefont{Chatrchyan}}
  \bibnamefont{et~al.} (\bibinfo{collaboration}{CMS}),
  \bibinfo{journal}{CMS-PAS-HIG-12-020}  (\bibinfo{year}{2012}{\natexlab{c}}).

\end{thebibliography}

\end{document}